\documentclass[usenatbib,usegraphicx]{mn2e}
\usepackage{amssymb}
\usepackage{bm}
\usepackage{color}
\usepackage{graphicx}
\usepackage[normalem]{ulem}

\newcommand{\halfspace}{\hspace{1pt}}

\newcommand{\Lya}{Ly$\alpha$}
\newcommand{\Lyb}{Ly$\beta$}
\newcommand\HI{{\hbox{H\halfspace$\rm \scriptstyle I$}}}

\newcommand\HeII{{\hbox{He\halfspace$\rm \scriptstyle II$}}}
\newcommand\HeIII{{\hbox{He\halfspace$\rm \scriptstyle III$}}}

\newcommand\lsim{~\lower.5ex\hbox{$\buildrel < \over \sim$}~}
\newcommand\gsim{~\lower.5ex\hbox{$\buildrel > \over \sim$}~}
\title[Helium reionization and the \Lya\ forest]{The effect of helium reionization on the \Lya\ forest hydrogen flux statistics}
       
\author[A. Meiksin, E. Puchwein]{
        Avery Meiksin$^{1}$\thanks{E-mail:\ meiksin@ed.ac.uk (AM)},
        Ewald Puchwein$^{2}$\\
        $^{1}$SUPA\thanks{Scottish Universities Physics Alliance},
	Institute for Astronomy, University of Edinburgh,
        Blackford Hill, Edinburgh\ EH9\ 3HJ, UK\\
        $^{2}$Leibniz-Institut f\"ur Astrophysik Potsdam (AIP), An der Sternwarte 16, 14482 Potsdam, Germany
}

\shortcites{}

\begin{document}

\date{Accepted 23 October 2024. Received 22 October 2024; in original form 19 July 2024.}
\pagerange{\pageref{firstpage}--\pageref{lastpage}} \pubyear{2024}
\maketitle
\label{firstpage}

\begin{abstract}
 %Max 250 words
We assess the impact of QSOs on the high redshift ($z>4$) Intergalactic Medium using Monte Carlo realisations of QSO populations and the \HeIII\ regions they generate, applied to the Sherwood-Relics simulations, allowing for uncertainties in the QSO luminosity function, its evolution, and QSO spectra and ages. While QSO luminosity functions based on optical-infrared selection are unable to reproduce the broadening \HI\ \Lya\ optical depth distributions at $z>5$, much broader distributions are found for the higher numbers of QSOs based on x-ray selection, suggesting a large QSO contribution to the ultra-violet background at $z>5$ may offer an alternative to late reionization models to account for the broad \HI\ \Lya\ optical depth distributions. Realisations using QSOs based on the higher QSO counts also much better recover the measured pixel flux auto-correlation function at $z>5$. The \HeIII\ regions from QSO sources according to both types of luminosity function suppress the pixel flux power spectrum on small scales, $k>0.02\,\mathrm{s\,km^{-1}}$, while enhancing it on larger, both by amounts of up to tens of percent at $z>4$, with the magnitude increasing with the intergalactic \HeIII\ filling factor and the boost in temperature within the \HeIII\ regions.
\end{abstract}


\section{Introduction}
\label{sec:Intro}

The large-scale structure of the Intergalactic Medium (IGM) is a well-established consequence of structure formation in a Cold Dark Matter (CDM) dominated cosmology \citep{2009RvMP...81.1405M}. Allowing for the completion of hydrogen reionization by $z=5$, the model accounts for the absorption properties of the IGM as measured in the spectra of high redshift Quasi-Stellar Objects (QSOs) \citep{2016ARA&A..54..313M}. On sub-megaparsec scales, however, there remain some discrepancies. While the inferred temperatures are consistent with hydrogen photoionization at $z>5$, detailed analyses find that the measured absorption line widths at moderate redshifts \citep{Theuns98, 1999ApJ...517...13B, Schaye00, MBM01} and other statistics \citep{2000MNRAS.314..566M, 2000MNRAS.317..989T,  2011MNRAS.410.1096B, 2012MNRAS.424.1723G} suggest an additional source of thermal energy is required at $2<z<5$, such as may be provided by \HeII\ reionization by QSOs \citep{2000ApJ...532..118M, 2000MNRAS.315..600T, BMW04, 2007MNRAS.380.1369T, 2009ApJ...694..842M}.

Another consequence of the additional heat input following \HeII\ reionization is the small-scale suppression of the flux power spectrum \citep{2013PhRvD..88d3502V, 2017PhLB..773..258G, 2017PhRvD..96b3522I, 2021MNRAS.506.4389G}. The uncertainty in the amount of suppression is a limiting factor in using the \Lya\ flux power spectrum as a test of alternative models to the $\Lambda$CDM model that affect structures on megaparsec scales such as Warm Dark Matter (WDM) models, alongside uncertainties in the peculiar velocity field and cosmic reionization \citep{2023MNRAS.519.6162P, 2023PhRvD.108b3502V, 2024PhRvD.109d3511I}.

While previous work accounted for the uncertain IGM temperature resulting from \HeII\ reionization through global approximations, such as by parametrizing the IGM equation of state as a polytrope \citep{2013PhRvD..88d3502V, 2018PhRvD..98h3540M}, or by modifying the heating rate either directly \citep{2000ApJ...534...57B, 2017MNRAS.464..897B} or indirectly through radiative transfer \citep{2007MNRAS.380.1369T, 2009ApJ...694..842M, 2020MNRAS.491.1736K, 2024MNRAS.533.2843A}, we estimate the impact of \HeII\ reionization by modelling the influence of \HeIII\ regions around individual QSO sources. The approach is similar to previous simulations including QSOs \citep{2009ApJ...694..842M, 2017MNRAS.465.3429C, 2020MNRAS.496.4372U}, but differs in several respects:\ 1. QSOs are distributed up to scales approaching the cosmic horizon because rare bright QSOs dominate the QSO contribution to fluctuations in the ultra-violet background (UVB) \citep{2020MNRAS.491.4884M}. 2. While \citet{2017MNRAS.465.3429C} ran simulations in large volumes, they did not allow for the elevated temperatures in the \HeIII\ regions of individual QSOs. 3. We allow for evolution in the QSO luminosity function, necessary because of the longevity of hot \HeIII\ regions. 4. We allow for finite QSO lifetimes. 5. We perform a separate random realization of the QSO luminosity function for each line of sight drawn from the simulation volume, ensuring each line of sight samples a statistically independent field of QSOs. As we show, the QSOs induce large-scale correlations in the radiation and temperature field, so that relatively few statistically independent lines of sight could be drawn from a single realization.

Several systematic uncertainties further complicate the inclusion of individual QSO sources in numerical simulations. QSOs are expected to have short duty cycles, corresponding to lifetimes of $10-1000$~Myr \citep{2013ApJ...762...70C, 2015MNRAS...453...2779E, 2024MNRAS.528.4466P}, although possibly flickering on timescales shorter than 1~Myr \citep{2021ApJ...917...38E}. The spectra of QSOs over frequencies affecting hydrogen and helium ionization are uncertain, with measurements for a mean power-law ($L_\nu\sim\nu^{-\alpha_Q}$) index shortwards of the \HI\ photoelectric edge of $\alpha_Q=0.56^{+{0.28}}_{-0.38}$ for nearby dim QSOs \citep{2004ApJ...615..135S} and a softer index $\alpha_Q=1.57\pm0.17$ for bright radio-quiet QSOs at moderate redshifts \citep{2002ApJ...565..773T}. The spectrum immediately shortward of the \HeII\ photoelectric edge is largely unconstrained by direct observations. AGN spectral models suggest the existence of a Big Bump, with $\nu L_\nu$ peaking at $\sim30$~eV, near the \HeII\ edge \citep{1987ApJ...323..456M, 2014MNRAS.437.2376R}, corresponding to the continuation of a hard spectrum shortward of the hydrogen edge, although these models may not apply to bright QSO sources. A comparison between \HI\ and \HeII\ \Lya\ forests at $2<z<3.5$ suggests bright QSOs are soft shortward of the \HeII\ photoelectric edge, with $\alpha_Q<-1.6$ \citep{2017MNRAS.471..255K}. Other spectral indices have been inferred in the literature, such as $\alpha_Q=1.7$ \citep{2015MNRAS.449.4204L} and $\alpha_Q=1.96$ \citep{1997ApJ...475..469Z}. The luminosity function of high redshift QSOs is also uncertain, with x-ray selected samples \citep{2019ApJ...884...19G} suggesting counts several times higher than optical-infrared selected samples \citep{2019MNRAS.488.1035K}, although some recent optical-infrared selected samples suggest numbers consistent with x-ray selected samples \citep{2023ApJ...955...60G, 2023ApJ...959...39H, 2023arXiv230811609F, 2023arXiv230801230M},
even though those discovered with the {\it James Webb Space Telescope} are  underluminous in x-rays \citep{2024arXiv240500504M}. The higher numbers of QSOs suggest they may provide a substantial contribution to hydrogen reionization and subsequent photoionization \citep{2023ApJ...955...60G, 2023arXiv230811609F, 2024ApJ...971...75M}. Numerical simulations that surveyed the range of impact on the IGM of all these uncertainties would be prohibititvely expensive.

Using the QSO luminosity functions from \citet{2019MNRAS.488.1035K} and \citet{2019ApJ...884...19G}, we derive the \HeIII\ regions from the ages and luminosities of individual QSO sources. We consider a range in QSO lifetimes and spectral hardnesses. Our main goal is to provide estimates of the range of impact of these QSO systematics on some of the key IGM statistics affected by QSOs at redshifts $z>4$, before substantial overlap of \HeIII\ regions so that the regions may still be treated individually. We focus on the 1-point and 2-point flux statistics, as measured by the effective optical depth distribution, the flux 1D auto-correlation function and the 1D flux power spectrum. We utilise the Sherwood-Relics suite of simulations \citep{2023MNRAS.519.6162P} to extract these statistics by embedding the simulated boxes in a field of randomly generated QSOs, including their \HeIII\ regions. The Sherwood-Relics simulations cover a range of simulation volumes, with side lengths from $10-160h^{-1}$~Mpc (comoving), and mass resolutions, with $512^3-2048^3$ initial gas particles per volume, allowing us to assess convergence with box size and resolution of the statistics examined.

We do not account for spatial correlations of the QSO sources. Doing so would require relating the dark matter halos in the simulations to QSOs. While there exist prescriptions for doing so \citep[e.g.][and references therein]{2013ApJ...762...70C, 2024MNRAS.528.4466P}, the dynamic range in halo masses in the Sherwood-Relics simulations is too small to adequately sample the bright end of the QSO luminosity function, which dominates the UVB fluctuations \citep{2020MNRAS.491.4884M}. The models are also constrained primarily by QSO clustering data at $z\lsim4$, while we address the impact of QSOs on the IGM at higher redshifts. The models have also been constructed for the lower QSO counts from optical-infrared selected samples, while we show here that only the higher counts from x-ray selected samples (and similar) have much effect on the \Lya\ optical depth distribution at $z>5$. Lastly, 
analytic estimates show that the contribution of spatial correlations of the QSOs to the UVB spatial correlations is small on the scales of interest compared with the fluctuations arising from shot noise \citep{2019MNRAS.482.4777M}. Realistic inclusion of the role of QSO clustering, which is possibly not negligible, on the results presented here must await additional clustering data at $z>4$ and future QSO population models.

This paper is organised as follows. The next section describes the method used. Sec.~\ref{sec:results}  presents results for the IGM properties and observational statistics. In Sec.~\ref{sec:disc} we discuss the results, and summarise the conclusions in Sec.~\ref{sec:conc}. Descriptions of several convergence tests are provided in the Appendix. Unless mentioned otherwise, we use the same $\Lambda CDM$ cosmological parameters as for the Sherwood-Relics simulations, based on \citet{2014A&A...571A..16P}:\ $\Omega_m=0.306$, $\Omega_\Lambda=0.692$, $\Omega_b=0.0482$, $h = 0.678$, $\sigma_8=0.829$ and $n_s = 0.961$, and refer to comoving and proper distance units with the prefix \lq c\rq\ and no prefix, respectively. 

\section{Method}
\label{sec:method}

\subsection{Monte Carlo simulation}
\label{subsec:mc}

A Monte Carlo approach is used to model the contribution of QSO sources to the photoionizing UVB. Two effects are simulated, the QSO discreteness (shot noise) contribution to the \HI-ionization rate and the heating arising from \HeII-photoionization. To account for the evolution of the QSO luminosity function, nested cubical boxes are constructed around the simulation volume. Each box is triple in length the previous box in the series, and centred on the centre of the original simulation volume. The final box is the largest which may be confined by the cosmic horizon of the simulation volume, or by 30 times the mean free path of hydrogen-ionizing photons, or having a radiative cooling time that is shorter than the age of the \HeIII\ region at the simulation redshift, whichever is smaller. (For the simulation redshifts and QSO lifetimes of interest, the latter corresponds to excluding boxes at $z>7$, with a negligible effect on the QSO contribution to the hydrogen-ionizing UVB.) Each successive box corresponds to successively earlier cosmic times according to the incremental redshift factor across the box. The space-time boundary of the nested boxes thus encloses the past light cone of the simulation volume.

The Sherwood-Relics suite is generated from cosmological hydrodynamical simulations using modified versions of the \texttt{P-GADGET3} code. The code uses a Tree-PM gravity solver for the gravitational interactions, and solves the fluid equations using a smoothed-particle hydrodynamics scheme. Further details are provided in \citet{2023MNRAS.519.6162P}. The simulations in the suite provide 5000 lines of sight through the simulation volume at each redshift dump. A separate Monte Carlo realization of the QSO distribution is drawn for each line of sight to avoid artificial correlations in the statistics along nearby lines of sight arising from shared UVB and thermal inhomogeneities from \HeIII\ regions when averaging the results. We use only models with homogeneous hydrogen reionization, completing at $z_r=6.0$. Although some inhomogeneous reionization models are available, our intention here is to isolate the effects UVB and thermal inhomogeneities arising from QSO sources have on the IGM.

The UVB contribution from the QSOs is computed by the following steps:\ (1) A random number of QSO sources is drawn in each box according to the expected number based on the QSO luminosity function within the absolute magnitude interval sampled. (This is fixed to the expected number if it exceeds $3\times10^4$, since the mean Poisson fluctuations in the number of sources is then below about a percent.)  (2)\ The QSO luminosities are sampled randomly from the QSO luminosity function. (3)\ The QSO sources are distributed randomly throughout the Monte Carlo QSO volume. (4)\ The contribution of the QSOs to the hydrogen ionization rate $\Gamma_Q(\mathbf{r})$ in a cell at comoving position $\mathbf{r}$ in the IGM simulation volume is computed from
\begin{equation}
\Gamma_Q(\mathbf{r}) =\frac{1}{4\pi} \frac{\sigma_L}{h_P(3+\alpha_Q)}{\sum_i}L_i\frac{e^{-\vert\mathbf{r}_i-\mathbf{r}\vert/\lambda_\mathrm{mfp}}}{\vert\mathbf{r}_i-\mathbf{r}\vert^2}
\label{eq:GammaQ}
\end{equation}
where $L_i$ is the UV luminosity at the hydrogen Lyman edge of source
$i$ at comoving position $\mathbf{r}_i$, and all sources are assumed to have
the same spectral index $\alpha_Q$, with $L_\nu\sim\nu^{-\alpha_Q}$, and to see the same comoving mean free path
$\lambda_\mathrm{mfp}$, fixed at the redshift of the box containing the QSO when born. The mean free path is adopted from \citet{2014MNRAS.445.1745W} for $z<5$ and \citet{2021MNRAS.508.1853B} for $z>5$. (From $z=4.6$ to 5.6 to 6, the comoving mean free path evolves from 79 to 16 to 7$h^{-1}$~cMpc, respectively. Using the mean free path at the redshift of the simulation volume is found to produce a negligible change in results.) We assume an escape fraction of unity for the ionizing radiation, consistent with QSOs at $z<1.5$ \citep{2014ApJ...794...75S}, while bright QSOs at $3.6<z<4.0$ have an average escape fraction of $\sim0.8$, with wide dispersion \citep{2016MNRAS.462.2478C}. For simplicity, we adopt a value of unity for all QSOs, noting that this provides an upper limit to the effect of QSOs for a given QSO luminosity function, but is a variable that should be considered along with others to cover a more complete range in the possible influence of QSOs on the IGM.

In Eq.~(\ref{eq:GammaQ}), we generally adopt $\alpha_Q=1.57$ \citep{2002ApJ...579..500T}, although we also consider models with a harder spectrum. For the latter, for simplicity we continue the power-law $\alpha_Q=0.61$ used by \citet{2019MNRAS.488.1035K} between 1450A and 912A to shorter wavelengths, as this is consistent with the determination of \citet{2004ApJ...615..135S}. In fact the spectra will be filtered in frequency by the IGM, as higher energy photons travel further than lower energy. The exponential in Eq.~(\ref{eq:GammaQ}) more generally would include a frequency-dependent factor.  For a diffuse IGM, with the frequency dependence of the IGM optical depth $\sim\nu^{-3}$, the metagalactic radiation from sources with $\alpha_Q=1.57$ hardens to $\nu^{-\alpha_\mathrm{eff}}$ with $\alpha_\mathrm{eff}\simeq0.25$, boosting the ionization rate by $\sim50$ percent compared with $\alpha_Q=1.57$.  At moderate to high redshifts, the optical depth is instead expected to be dominated by diffuse \HI\ absorption systems, giving a frequency dependence $\nu^{-3(\beta-1)}$, where $\beta$ describes the column density distribution of \HI\ absorption systems, $dN/dN_\mathrm{HI}\sim N_\mathrm{HI}^{-\beta}$ \citep{1996ApJ...461...20H, 2012ApJ...746..125H}. The IGM opacity of \citet{2012ApJ...746..125H} at moderate redshifts ($z\sim2-3$) is well-reproduced for $\beta=1.2$  \citep{2019MNRAS.482.4777M}, for which QSO sources with $\alpha_Q=1.57$ give a UVB frequency dependence $\alpha_\mathrm{eff}\simeq1.1$. By $z>4$, the IGM is sufficiently optically thick to ionizing radiation that the UVB increasingly reflects the spectrum
of local sources \citep[see figure~5 of][]{1996ApJ...461...20H}. It is shown in the Appendix this negligibly affects the distribution of \Lya\ optical depths. For simplicity, we ignore the frequency-dependence in the photoelectric optical depth.

QSOs with a finite lifetime $\tau_Q$ are accounted for as follows. The sources are assumed to have a constant birthrate per unit volume $\psi_Q(L)=\Phi_{Q, \mathrm{obs}}(L)/\tau_Q$, where $\Phi_Q(L)$ is the measured QSO luminosity function. The expected number of new QSOs introduced into a box of volume $V_\mathrm{box}$ is taken as $\Delta N_Q = V_\mathrm{box}\int dL\,\psi_Q \Delta t$, where the integral is carried over a given range in QSO luminosities and $\Delta t$ is the difference in cosmic time between the current box and the previous box in the hierarchy of nested boxes. (For the simulation volume, $\Delta t$ is taken to be half the light-crossing time of the box.) Ages are assigned to the new QSOs by uniformly sampling between 0 and $\Delta t$. The contribution from a QSO at comoving position $\mathbf{r_i}$
%and of age $t_{Q, i}$
to $\Gamma_Q$ in a cell at comoving position $\mathbf{r}$ within the simulation volume is counted in Eq.(\ref{eq:GammaQ})
% only if $ct_{Q,i} < \vert \mathbf{r_i}-\mathbf{r}\vert < c(t_{Q,i}+\tau_Q)$, where $c$ is the speed of light, to enforce causality.
only if $\Delta r_{i, \mathrm{min}} < \vert \mathbf{r_i}-\mathbf{r}\vert < \Delta r_{i, \mathrm{max}}$, where $\Delta r_{i, \mathrm{min}}$ and $\Delta r_{i, \mathrm{max}}$ are the minimum and maximum comoving distances, respectively, light from the QSO could travel given its redshift, age and lifetime, to enforce causality. This allows an early distant QSO to contribute to the UVB in the simulation volume even after the QSO has died.

A young QSO source $i$ of age $t_{Q,i}$ at the redshift of interest, will be surrounded by an expanding $\HeIII$ region for $t_{Q,i}<\tau_Q$. The (comoving) radius $R_{\mathrm{I}, i}$ of the \HeIII\ region increases following the causal formulation of \citet{2003AJ....126....1W}. For the QSO luminosity function of \citet{2019MNRAS.488.1035K}, the filling factor of the \HeIII\ regions is $\sim0.1$ at $z=4.6$, so that the neglect of overlapping \HeIII\ regions will still be a good approximation. For cells with $\vert \mathbf{r}_i-\mathbf{r}\vert<R_{\mathrm{I},i}$, the temperature of the gas in the cell at $\mathbf{r}$ is boosted by an amount $\Delta T_b$. Temperature boosts of $1-2\times10^4$~K are expected following \HeII\ reionization in the IGM for a soft spectrum QSO \citep{2007MNRAS.380.1369T, 2010MNRAS.401...77M, 2012MNRAS.423....7M},
although values of $\sim3\times10^4$~K may be reached near young QSOs with lifetimes shorter than 100~Myr or with hard spectra \citep{2009MNRAS.395..736B, 2023MNRAS.519.5743L}. We consider models with $0\le\Delta T_b\le 4\times10^4$~K. The temperature boost is assumed to vanish for sources with $t_{Q,i}-\tau_Q$ exceeding the cooling time of the gas. This ensures relic \HeIII\ regions are extinguished if they have not been sourced by an active QSO for such a long time that the \HeIII\ region would have cooled and recombined back to \HeII\ by the redshift of the simulation volume of interest.

Using the new photoionization and temperature fields, equilibrium \HI\ fractions are regenerated along the lines-of-sight drawn from the Sherwood-Relics simulations. The \Lya\ forest spectra are constructed allowing for a Voigt profile. Self-shielding from the radiation field in dense regions is found to negligibly affect the statistics discussed here, as described in the Appendix, and so is not generally included.

\subsection{Sources}
\label{subsec:sources}

QSO sources are drawn from the luminosity functions of \citet{2019MNRAS.488.1035K} (KWH19) for $4<z<6$ and \citet{2019ApJ...884...19G} (G19) for $5<z<6$. \citet{2019MNRAS.488.1035K} present three models. We adopt primarily their Model 1, but also consider the other models for comparison. We adopt Model 4 from G19; Model 3 gives very similar results. Because G19 provide no continuous redshift evolution between $z=4$ to $z=6$, we do not consider their lower redshift ($4<z<5$) models (Models 1 and 2). When adopting Model 4 of G19, we assume the QSO luminosity function is un-evolving to $z=7$. While this is perhaps an unwarranted extrapolation, it allows us to obtain an upper limit on the influence of QSO sources following the higher QSO counts in the G19 models; it is also consistent with the measurements of \citet{2023ApJ...959...39H} to $z\simeq7$. Unless stated otherwise, for both the KWH19 and G19 models we consider only QSOs brighter than $M_{1450}<-21$.

We generally assume galaxies and diffuse emission from Lyman Limits Systems \citep[e.g.][]{1996ApJ...461...20H}, give rise to a uniform UVB. It was pointed out, however, by \citet{2018MNRAS.473..560D} that this need not necessarily be so for the contribution from galaxies if they have high escape fractions $f_\mathrm{esc}$ of ionizing radiation at $z>5$. They argue that, if galaxies brighter than $M_\mathrm{AB, 1600}>-20.7$ provide the UVB at $z=5.6$, the resulting UVB fluctuations could produce the broad IGM optical depth distribution measured by \citet{2015MNRAS.447.3402B} for $f_\mathrm{esc}>0.5$.

\begin{figure*}
\scalebox{0.6}{\includegraphics{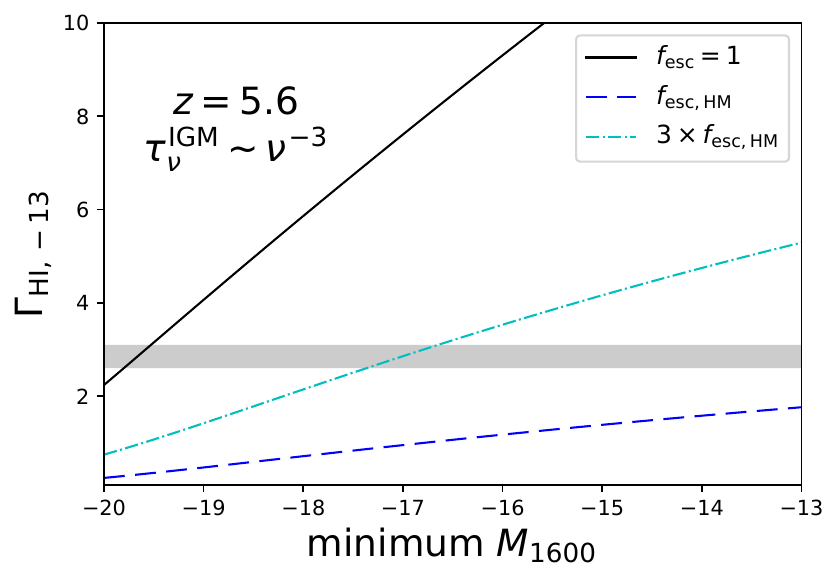}\includegraphics{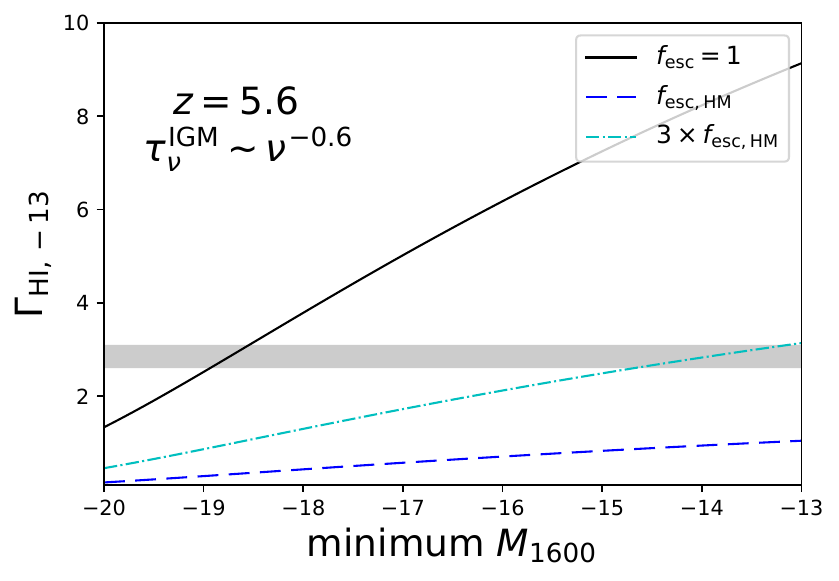}}
%\scalebox{0.45}{\includegraphics{fig01.pdf}}
%\vspace{-1.5cm}
\caption{The contribution of LBGs to the metagalactic \HI\ photoionization rate $\Gamma_\mathrm{HI}$ (in units of $10^{-13}\,\mathrm{s^{-1}\,atom^{-1}}$), for the indicated escape fractions, assuming an IGM optical depth at energies above the Lyman edge varying as $\nu^{-3}$ ({\it left-hand panel}) and $\nu^{-0.6}$ ({\it right-hand panel}). The grey shaded regions indicate the required level to reproduce the measured median \Lya\ optical depth of the IGM.
}
\label{fig:GamHI_LBG}
\end{figure*}

Based on more recent Lyman-Break Galaxy (LBG) counts \citep{2021AJ....162...47B} and measurements of the mean free path of intergalactic ionizing radiation \citep{2021MNRAS.508.1853B}, Fig.~\ref{fig:GamHI_LBG} shows that galaxies are unable to provide the ionizing backround required by the Sherwood-Relics simulations to match the median \Lya\ optical depth distribution measured by \citet{2022MNRAS.514...55B} at $z\simeq5.6$, assuming the escape fraction from \citet{2012ApJ...746..125H} ($f_\mathrm{esc}\simeq0.11$ at $z=5.6$), but may do so for an escape fraction three times higher ($f_\mathrm{esc}\simeq0.33$ at $z=5.6$), when also allowing for an extension of the luminosity function to $M_\mathrm{AB, 1600}\simeq-17.5$ (-14.5) for $\tau_\nu^\mathrm{IGM}\sim\nu^{-3}$ ($\tau_\nu^\mathrm{IGM}\sim\nu^{-0.6}$), comparable or somewhat fainter than the observed dimmest galaxies at $M_\mathrm{AB, 1600}<-16.5$. For such faint limits, the number of galaxies is sufficiently high to assure a smooth UVB contribution. For $f_\mathrm{esc}=1$, the UVB may be provided by LBGs with $M_\mathrm{AB, 1600}\lsim-19.5$ for diffuse IGM Lyman photon absorption ($\tau^\mathrm{IGM}_\nu\sim\nu^{-3}$) and $M_\mathrm{AB, 1600}\lsim-18.5$ for clumpy IGM Lyman photon absorption ($\tau^\mathrm{IGM}_\nu\sim\nu^{-0.6}$). We explore the case of a UVB provided by LBGs in Sec.~\ref{sec:disc} below.

\section{Results}
\label{sec:results}

\subsection{Structure of the IGM}
\label{subsec:IGMstruc}

\begin{figure}
%\scalebox{0.5}{\includegraphics{fig01_a.eps}\includegraphics{fig01_b.eps}}
\scalebox{0.45}{\includegraphics{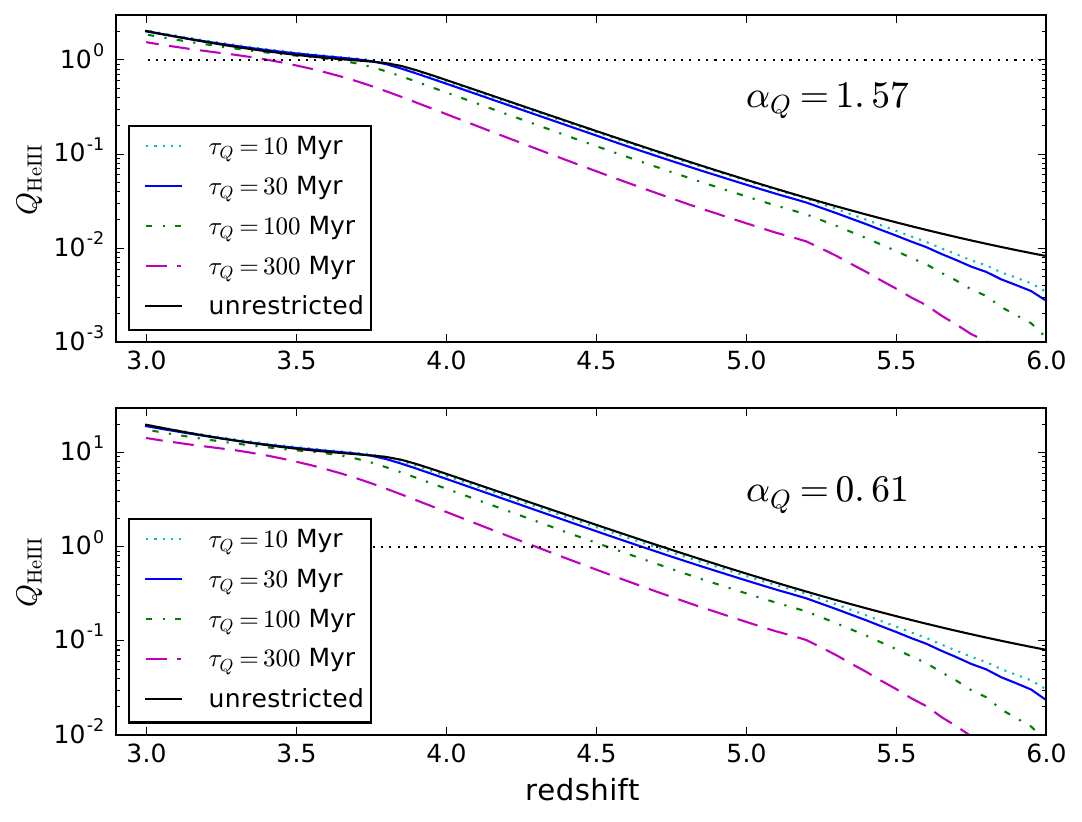}}
%\vspace{-1.5cm}
\caption{Evolution of the \HeIII\ porosity parameter for QSO luminosity function Model 1 of \citet{2019MNRAS.488.1035K} for a QSO spectral index shortward of the \HeII\ Lyman edge of $\alpha_Q=1.57$ ({\it upper panel}) and 0.61 ({\it lower panel}), for the indicated QSO lifetime $\tau_Q$. The black solid lines show the prediction assuming every ionizing photon reaching the ionization front is absorbed (using Eq.~[\ref{eq:Qev}] in the text). 
  }
\label{fig:QHeIII}
\end{figure}

The evolution of the \HeIII\ porosity parameter (the product of the \HeIII\ region volumes and their number per unit volume) for the KWH19 QSO luminosity function Model 1 is shown in Fig.~\ref{fig:QHeIII}. The QSO spectrum shortward of the \HeII\ Lyman edge is taken to be $\alpha_Q=1.57$ (upper panel) or 0.61 (lower panel). Assuming that every photon arriving at the ionization front is absorbed at the front gives the growth equation for porosity $Q$
\begin{equation}
\frac{dQ}{dt}=\frac{\dot n_S}{n} - \frac{Q}{\tau_\mathrm{rec}},
\label{eq:Qev}  
\end{equation}
for a combined source emissivity per volume $\dot n_S$, number density $n$ of the species being photoionized and recombination time $\tau_\mathrm{rec}$ \citep{1999ApJ...514..648M}. The results for photoionizing \HeII\ (assuming neutral helium was singly ionized by galaxies) are shown by the black curves in Fig.~\ref{fig:QHeIII}. The results are not very sensitive to $\tau_\mathrm{rec}$ because of the long recombination time, even allowing for a gas clumping factor up to $\sim3$.

For bright sources like QSOs, the assumption that all emitted photons are immediately absorbed expressed by Eq.~(\ref{eq:Qev}) requires the early part of the \HeIII\ front expansion to be superluminal. Enforcing causality, following \citet{2003AJ....126....1W}, restricts the expansion to being subluminal and results in the curves labelled by QSO lifetime $\tau_Q$ in Fig.\ref{fig:QHeIII}. The porosity parameter evolution is sensitive to the QSO lifetime because, for short-lived QSOs, the photons are soon completely absorbed, and the evolution follows Eq.~(\ref{eq:Qev}), while for longer lived QSOs, it takes longer for the ionization front to catch up to the predicted value assuming all emitted photons had been immediately absorbed, and so the evolution of the porosity lags behind Eq.(\ref{eq:Qev}). There is little difference in the porosity parameters for $\tau_Q<100$~Myr once $Q_\HeIII > 0.1$.

The porosity parameter corresponds to the filling factor of the ionized gas until it reaches unity. For a soft spectrum, $\alpha_Q=1.57$, \HeII\ reionization completes at $z\simeq3.4-3.7$, with the \HeIII\ porosity below 2 percent at $z\sim5.5$. For the harder spectrum, $\alpha_Q=0.61$, \HeII\ reionization completes much earlier, at $z\simeq4.3-4.7$, somewhat earlier than expected ($z\sim3$) based on estimates of the \HeII\ photoionization rate at $z<4$ \citep{2019ApJ...875..111W}. For KWH19 Model 3, \HeII\ reionization does not complete until as late as $z\simeq3.0$ for the softer spectrum. The hard spectrum case may be considered an upper limit to the \HeIII\ porosity for the KWH19 QSO luminosity functions. In contrast, the larger numbers of QSOs from the G19 QSO luminosity function give $Q_\HeIII\simeq0.2$ at $z\simeq5.6$ for $\alpha_Q=1.57$, and assuming the comoving luminosity function remains constant to higher redshifts. \citet{2024ApJ...971...75M} obtain a similar value; on interpolating to the KWH19 QSO luminosity function for $z<2$, their model predicts \HeII\ reionization ends at $z\gsim3$.

\begin{figure}
%\scalebox{0.5}{\includegraphics{fig01_a.eps}\includegraphics{fig01_b.eps}}
\scalebox{0.56}{\includegraphics{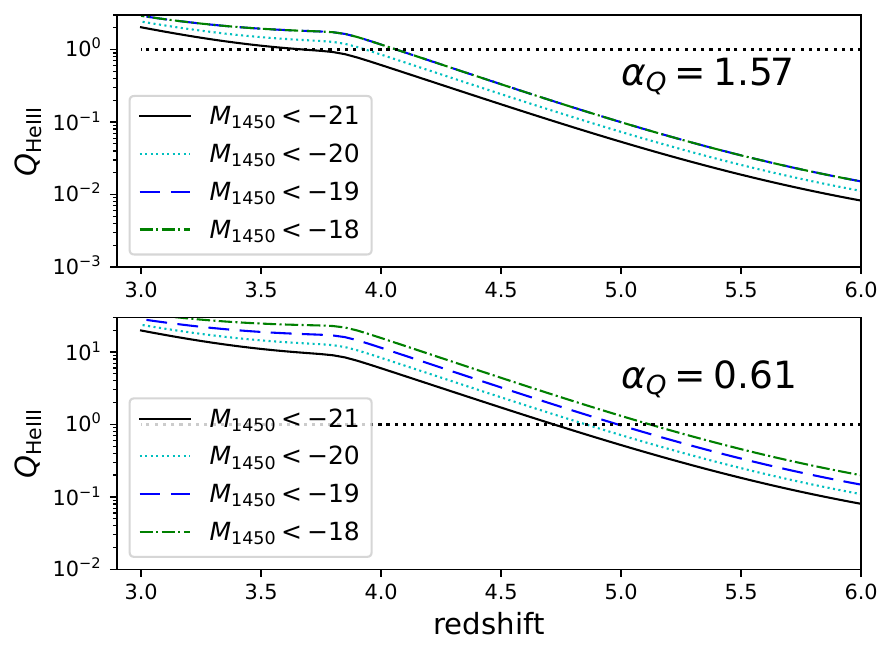}}
%\vspace{-1.5cm}
\caption{Evolution of the \HeIII\ porosity parameter for QSO luminosity function Model 1 of \citet{2019MNRAS.488.1035K} for the indicated QSO magnitude ranges. Shown for a QSO spectral index shortward of the \HeII\ Lyman edge of $\alpha_Q=1.57$ ({\it upper panel}) and 0.61 ({\it lower panel}). The curves show the predictions for the limiting case for which every ionizing photon reaching the ionization front is absorbed (using Eq.~[\ref{eq:Qev}] in the text).
  }
\label{fig:QHeIII_Mmax}
\end{figure}

Results from \emph{JWST} observations show the number of Type 1 AGN at $4<z<6$ for $-20<M_{1450}<-18$ exceeds the G19 QSO counts for these magnitudes by a factor of 10--20 \citep{2023ApJ...959...39H, 2023arXiv230801230M}. The extrapolation of the KWH19 Model 1 QSO luminosity function to these low luminosities approaches similar numbers over $4.1<z<4.7$. Follow-up \emph{Chandra} observations of the JADES survey AGN find that the vast majority of the sources do not emit X-rays, with upper limits a factor of several to two hundred below the expectation for optical/UV selected AGN \citep{2024arXiv240500504M}. It is unknown whether they emit \HeII-ionizing photons. Assuming QSOs with $M_{1450}>-21$ produce \HeII-ionizing radiation, Fig.~\ref{fig:QHeIII_Mmax} shows that a large \HeIII\ porosity parameter at $z=4.6$ may arise from the KWH19 Model 1 QSO luminosity function. Extrapolating to $M_{1450}<-18$ gives $Q_\HeIII\simeq0.35$ for $\alpha_Q=1.57$, compared with $Q_\HeIII\simeq0.14$ for $M_{1450}<-21$. For $M_{1450}<-18$ and $\alpha_Q=0.61$, $Q_\HeIII>1$ at $z<5.1$. Extrapolating to $M_{1450}<-18$ results in QSOs providing a third of the UV hydrogen-ionizing photons at $z=4.6$, at the rate $\Gamma_\mathrm{HI} \simeq 2\times10^{-13}\,\mathrm{s}^{-1}$, compared with a total rate $\Gamma_\mathrm{HI} \simeq 6\times10^{-13}\,\mathrm{s}^{-1}$ required to match the measured median effective \Lya\ optical depth for a uniform UVB model. By comparison, at $z=5.6$ the QSO counts for KWH19 Model 2 are highest extrapolated to $M_{1450}<-19$. Extrapolating to $M_{1450}<-18$ produces $\Gamma_\mathrm{HI} \simeq 1\times10^{-14}\,\mathrm{s}^{-1}$, compared with a total rate $\Gamma_\mathrm{HI} \simeq 2.5\times10^{-13}\,\mathrm{s}^{-1}$ required to match the measured median effective \Lya\ optical depth for a uniform UVB model, so that even extrapolating to $M_{1450}<-18$, the QSOs contribute negligibly to the UVB.

\begin{figure}
%\scalebox{0.5}{\includegraphics{fig01_a.eps}\includegraphics{fig01_b.eps}}
\scalebox{0.55}{\includegraphics{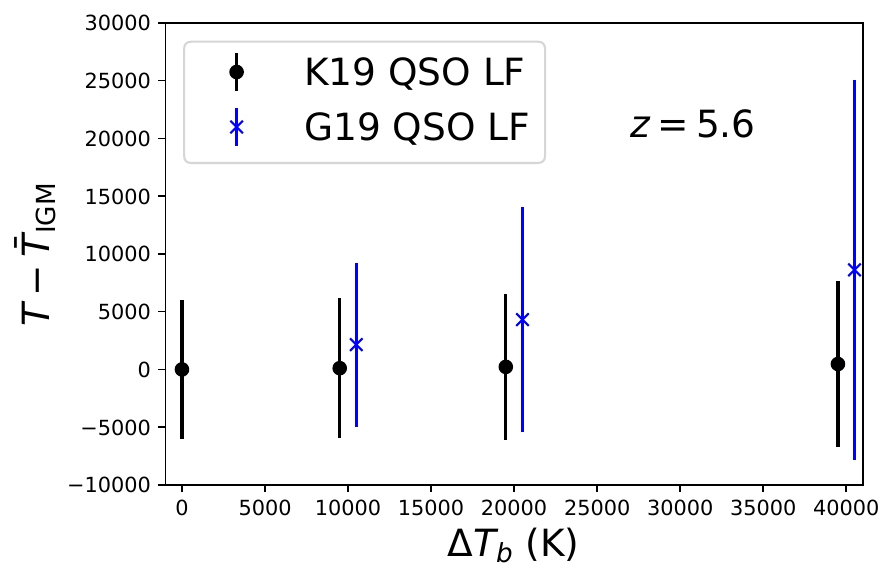}}
%\vspace{-1.5cm}
\caption{The mean difference in the IGM temperature and the mean IGM temperature $\bar T_\mathrm{IGM}$ in the original Sherwood-Relics simulation (dots and crosses), and the \emph{rms} spread (error bars), at $z=5.6$, for temperature boosts of $\Delta T_b=0$, $10^4$~K, $2\times10^4$~K and $4\times10^4$~K for QSO luminosity functions of \citet{2019MNRAS.488.1035K}, Model 1 (black circles), and \citet{2019ApJ...884...19G}, Model 4 (blue crosses). Both are for a QSO spectral index shortward of the \HeII\ Lyman edge of $\alpha_Q=1.57$. Negative values in the range arise from the temperature fluctuations in the original simulation. (The points are slightly offset in $\Delta T_b$ for visibility.) 
  }
\label{fig:dTmean}
\end{figure}

The mean difference between the IGM temperature and the mean IGM temperature $\bar T_\mathrm{IGM}$ in the original Sherwood-Relics simulation, and the \emph{rms} spread are shown in Fig.~\ref{fig:dTmean} at $z=5.6$ for the QSO luminosity functions of \citet{2019MNRAS.488.1035K} (Model 1), and \citet{2019ApJ...884...19G} (Model 4), for various temperature boosts $\Delta T_b$ inside \HeIII\ regions. Both are for a QSO spectral index shortward of the \HeII\ Lyman edge of $\alpha_Q=1.57$. For the KWH19 luminosity function, the IGM temperature increases and spread introduced by QSOs are moderate, corresponding to the low filling factor ($\sim0.01$) of the \HeIII\ regions. By contrast, for the G19 luminosity function, giving a \HeIII\ filling factor of $\sim0.2$, both the mean IGM temperature and its \emph{rms} spread increase with the temperature boost. The \emph{rms} spread includes the spread in temperature in the original Sherwood-Relics simulation.

\begin{figure*}
\scalebox{0.6}{\includegraphics{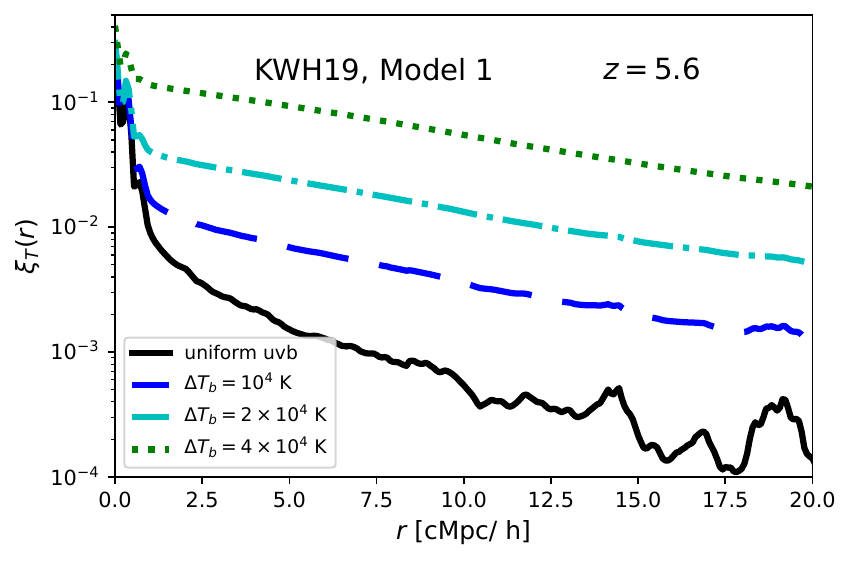}\includegraphics{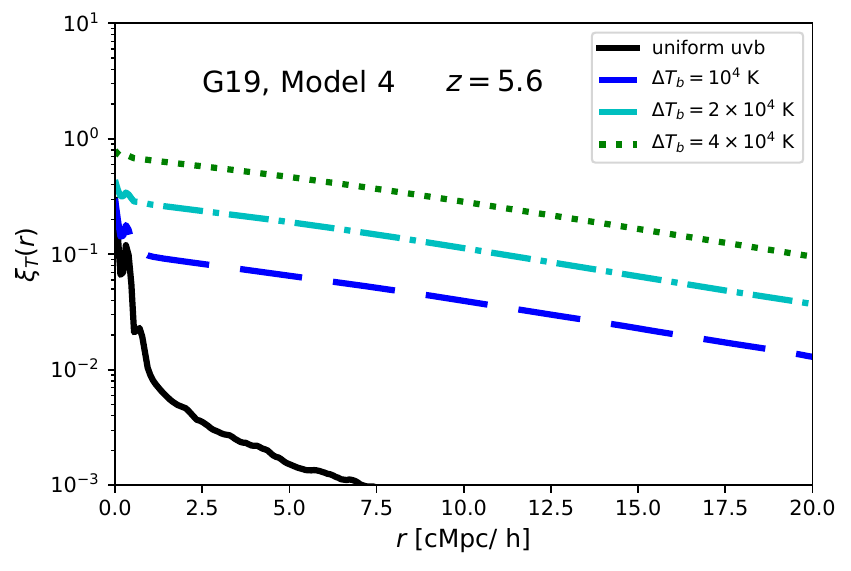}}
%\scalebox{0.5}{\includegraphics{fig02.pdf}}
%\vspace{-1.5cm}
\caption{The IGM temperature auto-correlation function at $z=5.6$ for temperature boosts of $\Delta T_b = 10^4$~K, $2\times10^4$~K and $4\times10^4$~K, compared with a uniform UVB, for QSO luminosity functions of \citet{2019MNRAS.488.1035K}, Model 1 ({\it left-hand panel}), and \citet{2019ApJ...884...19G}, Model 4 ({\it right-hand panel}). Both are for a QSO spectral index shortward of the \HeII\ Lyman edge of $\alpha_Q=1.57$. 
  }
\label{fig:xiT}
\end{figure*}

The temperature auto-correlation function at $z=5.6$ is shown in Fig.~\ref{fig:xiT}. The \HeIII\ regions boost the spatial correlations, which strengthen with $\Delta T_b$. The correlations are an order of magnitude stronger for the G19 QSO luminosity function compared with KWH19. For both cases, the correlations diminish approximately exponentially at large distances.

\begin{figure}
%\scalebox{0.5}{\includegraphics{fig03_a.pdf}\includegraphics{fig03_b.pdf}}
\scalebox{0.55}{\includegraphics{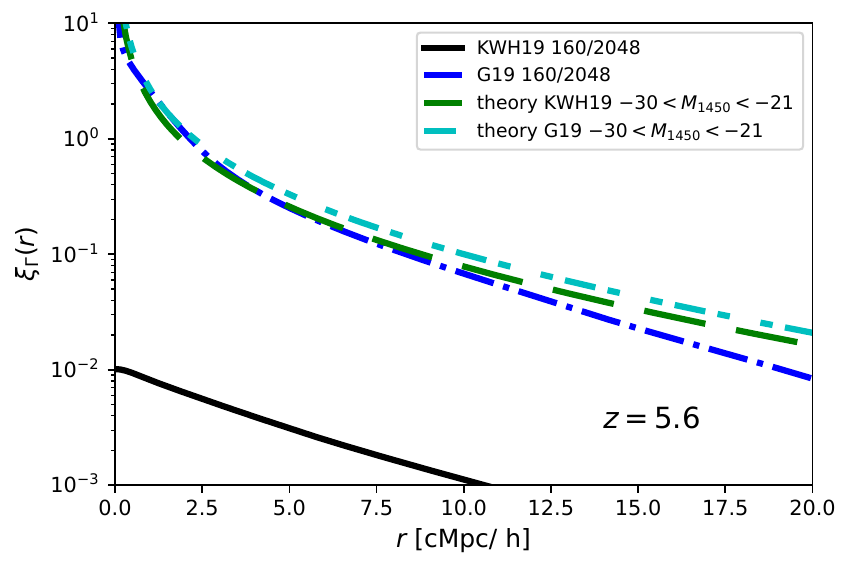}}
%\vspace{-1.5cm}
\caption{The UVB photoionization rate auto-correlation function at $z=5.6$ for QSO luminosity functions of \citet{2019MNRAS.488.1035K}, Model 1 (black solid line), and \citet{2019ApJ...884...19G}, Model 4 (blue dot-dashed line). Also shown are the theoretical expectation values. (See the Appendix for a discussion of the anomalously weak correlations for the KWH19 luminosity function.)
  }
\label{fig:xiG}
\end{figure}

The auto-correlation function of the photoionization rate from the metagalactic UVB, including both galaxies and QSOs, is shown in Fig.~\ref{fig:xiG} for the QSO luminosity functions KWH19 Model 1 (black solid line), and G19 Model 4 (blue dot-dashed line). Also shown are the theoretical predictions following \citet{1992MNRAS.258...45Z}. The correlations for the KWH19 luminosity function are much weaker than the theoretical expectation. As discussed in the Appendix, the theoretical formulation breaks down when the fluctuations are dominated by bright QSOs that are very rare within the horizon, especially when the luminosity function is evolving.

\subsection{Observational signatures}
\label{subsec:obssig}

The enhanced UVB spatial and temperature correlations give rise to an enhancement in \Lya\ transmission spatial correlations. We consider 1-pt and 2-pt statistics of the line-of-sight transmission.

\subsubsection{Effective optical depth distribution}
\label{subsubsec:taueff}
 
Measurements of QSO spectra show a broadening distribution of \Lya\ optical depth, averaged in $50h^{-1}$~Mpc comoving regions, with increasing redshift \citep{2015MNRAS.447.3402B, 2018ApJ...864...53E, 2019MNRAS.485L..24K, 2022MNRAS.514...55B}. Optical depth distributions were extracted from the simulations, including  QSO sources added from the KWH19 and G19 QSO luminosity functions. The \Lya\ forest spectra are drawn from the $160h^{-1}$~cMpc simulation with $2048^3$ initial gas particles. (See the Appendix for convergence tests with box size and gas particle resolution.)

\begin{figure}
%\scalebox{0.5}{\includegraphics{fig01_a.eps}\includegraphics{fig01_b.eps}}
\scalebox{0.55}{\includegraphics{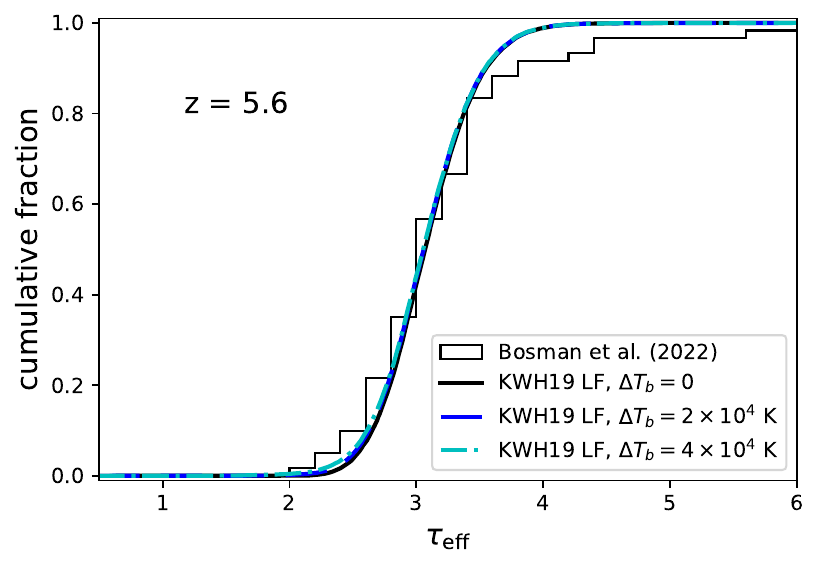}}
%\vspace{-1.5cm}
\caption{Distribution of effective optical depth averaged over regions $50h^{-1}$~cMpc across, shown for the KWH19 Model 1 luminosity function for a range in \HeIII\ region temperature boosts $\Delta T_b$. The result for $\Delta T_b=0$ is nearly identical to the result for a uniform UVB. Measurements from \citet{2022MNRAS.514...55B} are shown for comparison.
}
\label{fig:taueff_KWH19}
\end{figure}  
The optical depth distribution for the KWH19 Model 1 luminosity function at $z=5.6$ is shown in Fig.~\ref{fig:taueff_KWH19} for a variety of \HeIII\ region temperature boosts $\Delta T_b$. A QSO spectral index $\alpha_Q=1.57$ is assumed. The filling factor of \HeIII\ regions is $\sim0.01$. The distributions are insensitive to the boost temperature, and are all narrower than the distribution based on the measurements of \citet{2022MNRAS.514...55B} (using all absorption regions detected with $2\sigma$ confidence).

\begin{figure*}
\scalebox{0.6}{\includegraphics{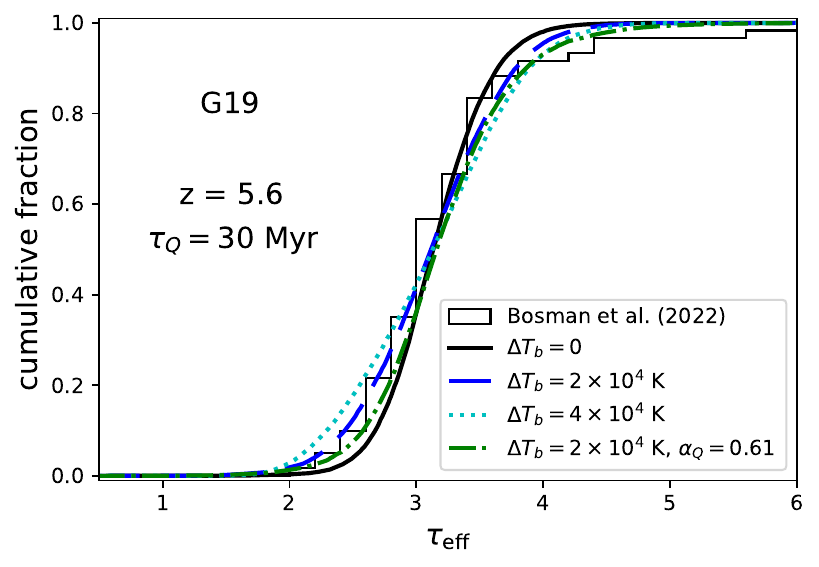}\includegraphics{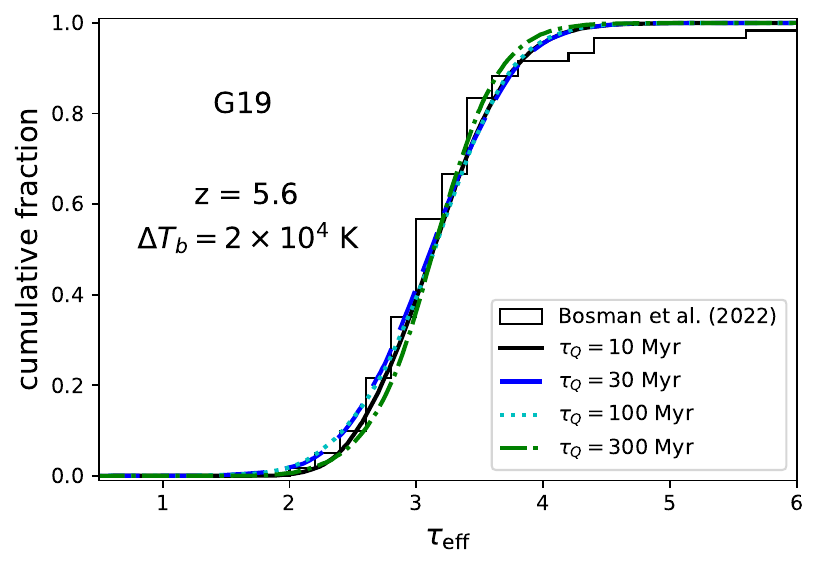}}
%\scalebox{0.5}{\includegraphics{fig06.pdf}}
%\vspace{-1.5cm}
\caption{Distribution of effective optical depth averaged over regions $50h^{-1}$~cMpc across, shown for the G19 Model 4 luminosity function. Measurements from  \citet{2022MNRAS.514...55B} are shown for comparison. ({\it Left-hand panel}):\ A QSO lifetime  $\tau_Q=30$~Myr is assumed. Increasing the temperature boost $\Delta T_b$, broadens the distribution (for QSO spectral index $\alpha_Q=1.57$), as does hardening the QSO spectrum to $\alpha_Q=0.61$. ({\it Right-hand panel}):\ A temperature boost $\Delta T_b=2\times10^4$~K is assumed. A QSO lifetime as great as 300~Myr results in a narrower distribution.
  }
\label{fig:taueff_G19}
\end{figure*}  
Increasing the temperature boost $\Delta T_b$ in the \HeIII\ regions broadens the distribution for the G19 QSO luminosity function (Model 4), as shown in the left-hand panel of Fig.~\ref{fig:taueff_G19}, with \HeIII\ filling factors of $\sim0.2$. The temperature boost $\Delta T_b=4\times10^4$~K may overly broaden the distrubtion. Hardening the spectrum to $\alpha_Q=0.61$ somewhat broadens the distribution at the high $\tau_\mathrm{eff}$ values, although the \HeIII\ filling factor has increased from 0.21 to 0.81, at which the approximation of treating the growth of the \HeIII\ regions as independent is expected to break down because of substantial overlap. Increasing the lifetime of the QSOs to as great as 300~Myr results in a narrower distribution than for shorter lifetimes, as shown in the right-hand panel. This is a consequence of a smaller \HeIII\ filling factor of $\sim0.1$, half that for the shorter QSO lifetime cases. The simulations still produce almost no regions with $\tau_\mathrm{eff}>5$, in contrast to the observations.

\begin{figure}
%\scalebox{0.5}{\includegraphics{fig06a.pdf}\includegraphics{fig06b.pdf}}
\scalebox{0.55}{\includegraphics{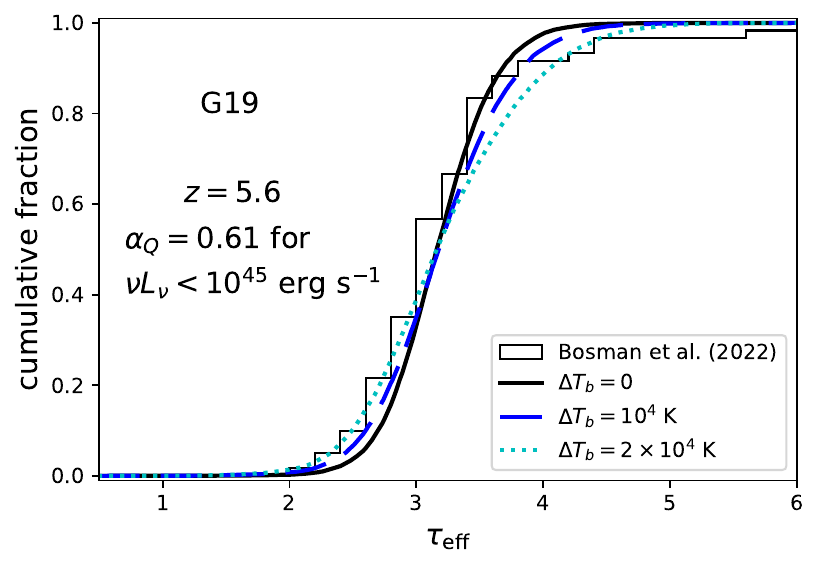}}
%\vspace{-1.5cm}
\caption{Distribution of effective optical depth averaged over regions $50h^{-1}$~cMpc across, shown for the G19 Model 4 luminosity function, with QSO spectral index $\alpha_Q=1.57$ (0.61) for QSOs with luminosity $\nu L_\nu$ at the Lyman edge greater (less) than $10^{45}\,\mathrm{erg\, s^{-1}}$, for a variety of \HeIII\ temperature boosts $\Delta T_b$. The QSO lifetime is $\tau_Q=30$~Myr. Measurements from  \citet{2022MNRAS.514...55B} are shown for comparison.
}
\label{fig:taueff_G19hy}
\end{figure} 
Allowing for harder lower luminosity sources, with $\alpha_Q=0.61$ for $\nu L_\nu<10^{45}\,\mathrm{erg\,s^{-1}}$ at the Lyman edge, consistent with \citet{2004ApJ...615..135S}, while retaining $\alpha_Q=1.57$ for higher luminosity sources, broadens the effective optical depth distribution at $z=5.6$ compared with taking $\alpha_Q=1.57$ for all sources, as shown in Fig.~\ref{fig:taueff_G19hy}. The \HeIII\ filling factor is 0.49. The distribution with $\Delta T_b=2\times10^4$~K better recovers the distribution than fixing $\alpha_Q=1.57$ or 0.61 for all sources (Fig.~\ref{fig:taueff_G19}).

\subsubsection{Pixel flux auto-correlation function}
\label{subsubsec:xiF}

\begin{figure*}
%\scalebox{0.5}{\includegraphics{fig01_a.eps}\includegraphics{fig01_b.eps}}
\scalebox{0.6}{\includegraphics{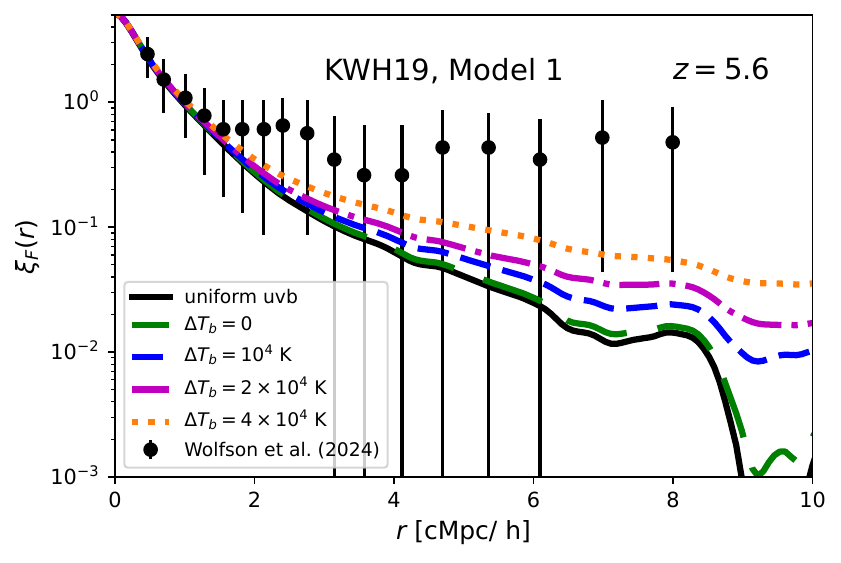}\includegraphics{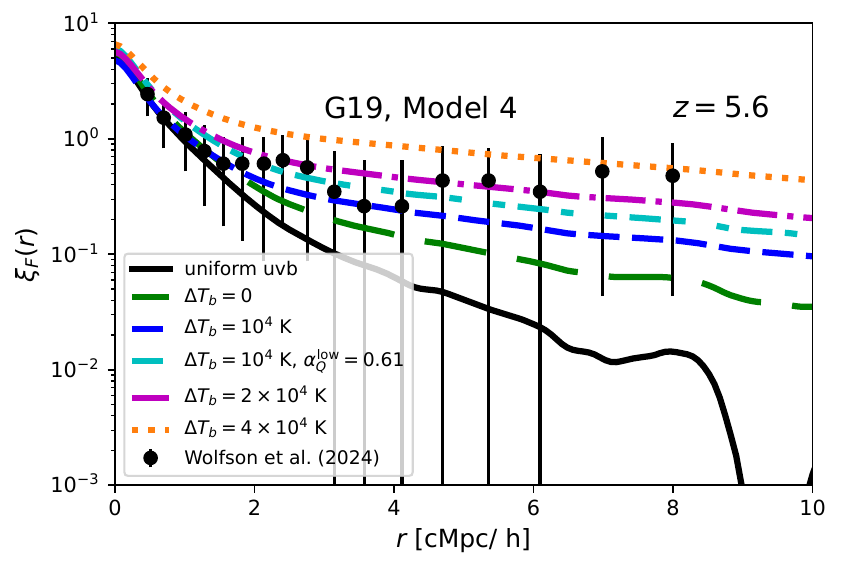}}
%\vspace{-1.5cm}
\caption{The pixel flux auto-correlation function, shown for the KWH19 ({\it left-hand panel}) and G19 ({\it right-hand panel}) QSO luminosity functions for a variety of \HeIII\ region temperature boosts $\Delta T_b$. Also shown in the right-hand panel is a case with a hard spectrum for low luminosity sources (see text). Measurements from \citet{2024MNRAS.531.3069W} are shown for comparison. Increasing the temperature boost strengthens the correlations at large separations.
}
\label{fig:xiF}
\end{figure*} 
The pixel flux auto-correlation function at $z=5.6$ for lag distance $r$, defined by $\xi_F(r)=\langle F(r_1)F(r_1+r)\rangle/\langle F(r_1)\rangle^2-1$ averaged over all $r_1$ along the lines of sight, where $F=\exp(-\tau)$ is the pixel flux relative to the QSO continuum ($F=1$), is shown in Fig.~\ref{fig:xiF} for the KWH19 (Model 1) and G19 (Model 4) QSO luminosity functions, both for a QSO spectral index $\alpha_Q=1.57$, and for a range of temperature boosts $\Delta T_b$. Also shown is a case for the G19 (Model 4) QSO luminosity function wth a harder spectrum ($\alpha_Q=0.61$) for sources with $\nu L_\nu<10^{45}\,\mathrm{erg\,s^{-1}}$. The results from the $160h^{-1}$~cMpc box are used.

The strength of the correlations decay nearly exponentially for small separations, and vanish at separations exceeding $8h^{-1}$~cMpc for a uniform UVB. As shown in the left-hand panel of Fig.~\ref{fig:xiF}, allowing for QSOs without a temperature boost in \HeIII\ regions only slightly enhances the correlations at large separations. The correlations at small separations are largely independent of any temperature boost, but are strengthened at large separations by an amount increasing with the magnitude of the temperature boost. The correlations for the KWH19 QSO luminosity function agree well with the measurements of \citet{2024MNRAS.531.3069W} for separations smaller than $1.5h^{-1}$~cMpc, but fall increasingly short of the measured values as the separation increases.

The predicted correlations for the G19 QSO luminosity function are similarly nearly independent of temperature boost and agree well with the measured strength for separations below $1.5h^{-1}$~cMpc. For temperature boosts of $\Delta T_b=1-2\times10^4$~K, the correlations match the values well over all scales measured, while for $\Delta T_b=4\times10^4$~K the predicted correlations somewhat exceed the measured values. Adopting a hard spectrum for lower luminosity sources enhances the correlations (shown in Fig.~\ref{fig:xiF} for $\Delta T_b=10^4$~K), with the amount of the enhancement diminishing with increasing $\Delta T_b$, so that results are nearly identical for $\Delta T_b=4\times10^4$~K (not shown).

\subsubsection{Pixel flux power spectrum}
\label{subsubsec:PkF}

\begin{figure*}
%\scalebox{0.5}{\includegraphics{fig01_a.eps}\includegraphics{fig01_b.eps}}
\scalebox{0.6}{\includegraphics{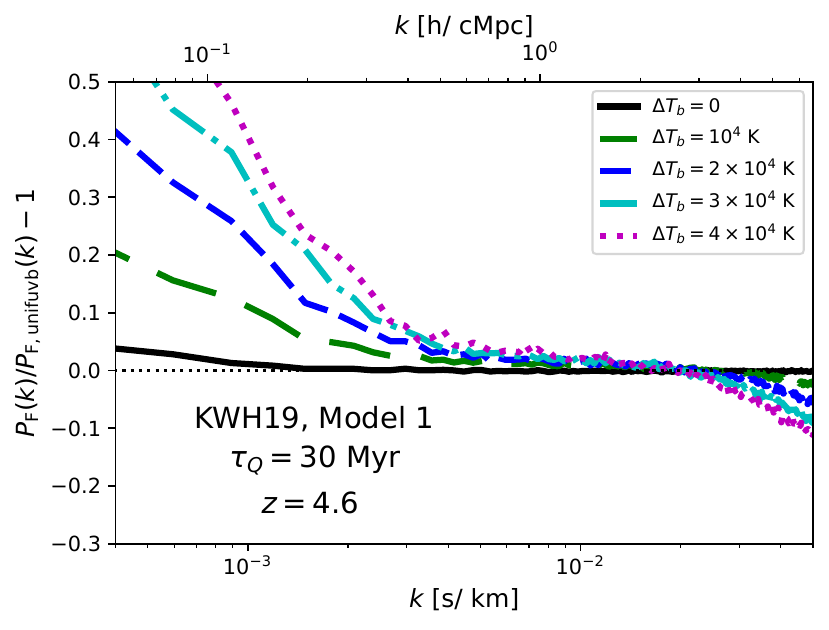}\includegraphics{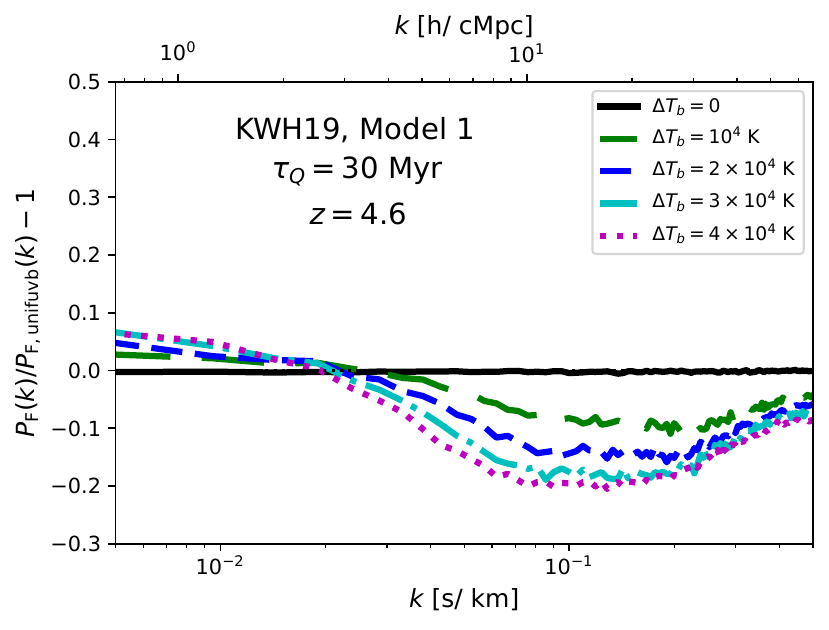}}
%\vspace{-1.5cm}
\caption{The relative difference in the pixel flux power spectrum at $z=4.6$ including QSOs following the KWH19 Model 1 QSO luminosity function, for the indicated \HeIII-region temperature boosts $\Delta T_b$ and with a QSO lifetime $\tau_Q=30$~Myr, compared with a homogeneous UVB. {\it Left-hand panel}:\ $160h^{-1}$~cMpc box. {\it Right-hand panel}:\ $10h^{-1}$~cMpc box.
}
\label{fig:dPkF_dTb}
\end{figure*} 

The relative change $\delta P_F(k)=P_F(k)/P_{F, \mathrm{unif uvb}}-1$, in the 1D pixel flux power spectrum at $z=4.6$ compared with a uniform UVB is shown in Fig.~\ref{fig:dPkF_dTb} for QSOs following the KWH19 Model 1 QSO luminosity function, for a range of \HeIII-region temperature boosts $\Delta T_b$. As discussed in the Appendix, convergence at high $k$ values is best achieved in a 10$h^{-1}$~cMpc box; this is used in the right-hand panel for $1024^3$ initial gas particles. For longer wavelength modes, we use results from a 160$h^{-1}$~cMpc box with initial gas particle number $2048^3$; the results are well-converged at small $k$, as discussed in the Appendix.

The left-hand panel shows a small rise in power of a few percent for long wavelength modes for $\Delta T_b=0$ (black solid curve), consistent with previous findings \citep{2004ApJ...610..642C, 2004MNRAS.350.1107M}. Allowing for temperature boosts in the \HeIII\ regions greatly magnifies the relative power spectrum, with $\delta P_F(k)$ exceeding 25 percent on scales $k<0.001\,\mathrm{s\,km^{-1}}$ for $\Delta T_b \ge 2\times10^4$~K. By contrast, power is suppressed at high wavenumbers by an amount increasing with $\Delta T_b$, reaching 20 percent suppression, as shown in the right-hand panel. In the region around $k\simeq0.02-0.03\,\mathrm{s\,km^{-1}}$, the power spectrum is unaffected; this region is fairly insensitive to $\Delta T_b$ over the range considered.

\begin{figure}
%\scalebox{0.5}{\includegraphics{fig06a.pdf}\includegraphics{fig06b.pdf}}
\scalebox{0.55}{\includegraphics{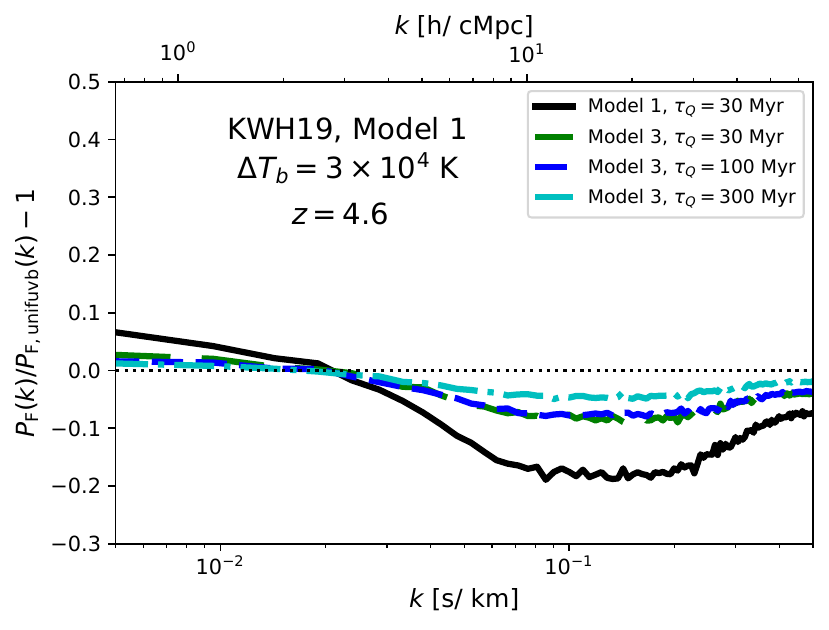}}
%\vspace{-1.5cm}
\caption{The relative change in the pixel flux power spectrum for different KWH19 QSO luminosity function models and QSO lifetimes, as indicated. Shown at $z=4.6$ for a \HeIII-region temperature boost $\Delta T_b=3\times10^4$~K.
}
\label{fig:dPkF_models}
\end{figure} 

\begin{figure}
%\scalebox{0. \HeIII-region temperature boost $\Delta T_b=2\times10^4$~K..  5}{\includegraphics{fig06a.pdf}\includegraphics{fig06b.pdf}}
\scalebox{0.55}{\includegraphics{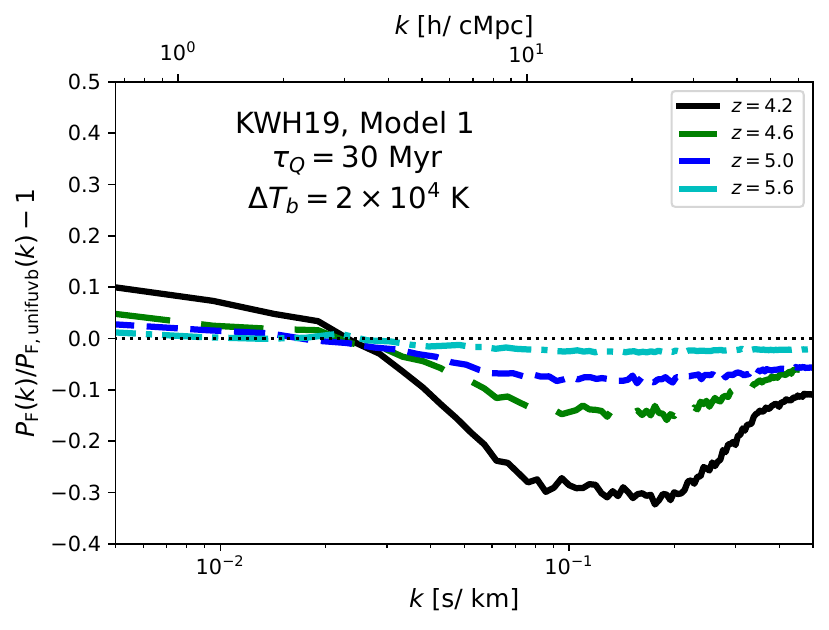}}
%\vspace{-1.5cm}
\caption{The evolution in the relative change in the pixel flux power spectrum for KWH19 Model 1 QSO luminosity function for QSO lifetime $\tau_Q=30$~Myr and \HeIII-region temperature boost $\Delta T_b=2\times10^4$~K.
}
\label{fig:dPkF_evol}
\end{figure} 

For a fixed temperature boost $\Delta T_b$, the amount of suppression in power at $k>0.02\,\mathrm{s\,km^{-1}}$ and enhancement at lower wavenumbers are also sensitive to the fraction of the IGM ionized to \HeIII, as shown in Figs.~\ref{fig:dPkF_models} and \ref{fig:dPkF_evol}. The suppression (enhancement) in power in Fig.~\ref{fig:dPkF_models} at high (low) $k$ is larger for Model 1 than for Model 3. For Model 1, the \HeIII\ filling factor at $z=4.6$ is 0.19 for $\tau_Q=30$~Myr; for Model 3, it's 0.085 for $\tau_Q=30$~Myr, 0.074 for $\tau_Q=100$~Myr and 0.043 for $\tau_Q=300$~Myr. The sensitivity of the amount of power suppression to the QSO lifetime is through the effect of QSO lifetime on the \HeIII\ filling factor (Fig.~\ref{fig:QHeIII}).

In Fig.~\ref{fig:dPkF_evol}, the suppression in power at $k>0.02\,\mathrm{s\,km^{-1}}$ for QSOs following the KWH19 Model 1 luminosity function increases with decreasing redshift, as the \HeIII\ filling factor increases. Over $z=5.6$, 5.0, 4.6 and 4.2, the respective values of the \HeIII\ filling factors are 0.015, 0.074, 0.18 and 0.45. Similarly, the enhancement in power at $k<0.02\,\mathrm{s\,km^{-1}}$ increases with increasing \HeIII\ filling factor.

\section{Discussion}
\label{sec:disc}

\subsection{\Lya\ optical depth distribution}
\label{subsec:taudist}

Allowing for photoionization of hydrogen and helium by discrete QSO sources induces correlations in the physical and observable properties of the hydrogen \Lya\ forest. We have examined the sensitivity of the \HI\ \Lya\ optical depth distribution, the pixel flux auto-correlation function and the pixel flux power spectrum to the luminosity function, spectra and lifetimes of the QSOs.

While previous simulations reproduced the measured \Lya\ optical depth distribution at redshifts $z<5$, a uniform UVB fails to recover the breadth of the distribution at higher redshifts \citep{2015MNRAS.447.3402B, 2018MNRAS.479.1055B, 2018ApJ...864...53E, 2020ApJ...904...26Y}, leading to the suggestion of large fluctuations in the UVB arising from fluctuations in the Lyman photon mean free path \citep{2016MNRAS.460.1328D, 2018MNRAS.473..560D}, large flucuations in the post-reionization IGM temperature \citep{2015ApJ...813L..38D}, or a large contribution of QSOs to the UVB \citep{2017MNRAS.465.3429C, 2020MNRAS.491.4884M}.

Post-reionization thermal fluctuations were not found adequate to account for the breadth of the optical depth distribution in the simulations of \citet{2018MNRAS.477.5501K}. Moreover, the regions around QSOs showing especially opaque regions in their spectra are under-dense in galaxies, inferred from an underdensity in Ly$\alpha$-emitting galaxies, so that they should have been reionzied later and so be warmer than the average IGM and so less opaque \citep{2021ApJ...923...87C}. The success of models including QSOs requires a higher QSO contribution to the UVB than given by the luminosity functions in either KWH19 or G19. An alternative possibility is that the high optical depth regions include neutral islands that persist to $z\simeq5.3$ \citep{2019MNRAS.485L..24K, 2020MNRAS.491.1736K, 2020MNRAS.494.3080N, 2022MNRAS.514...55B}.

A larger impact of QSOs on the optical depth distribution is found when allowing for the associated temperature boost in the \HeIII\ regions, an effect not included in \citet{2017MNRAS.465.3429C}. While after including the resulting thermal fluctuations in the IGM, the QSOs from the KWH19 luminosity functions are still found inadequate to account for the width of the optical depth distribution at $z\simeq5.6$, adopting instead the higher number of QSOs in the G19 luminosity function recovers most of the broadening. The additional temperature fluctuations allow for lower QSO counts than required by \citet{2017MNRAS.465.3429C}, as they anticipated. Recovering the fraction of regions with $\tau_\mathrm{eff}>5$ measured, however, is still difficult. Adopting harder spectra for lower luminosity QSOs further broadens the distribution. Including the expected fluctuations in the mean free path of Lyman continuum photons should also further broaden the distribution. As discussed in the Appendix, increasing the simulation box size beyond $160h^{-1}$~cMpc may also somewhat further increase the width of the distribution. Effects of source clustering are not included in these simulations. Clustered sources will produce larger patches of higher than average ionization, possibly shifting the median optical depth to lower values. This would require a smaller overall UVB to compensate, which may increase the number of high optical depth regions. This possibility is being explored elsewhere.

\begin{figure}
%\scalebox{0.5}{\includegraphics{fig01_a.eps}\includegraphics{fig01_b.eps}}
\scalebox{0.55}{\includegraphics{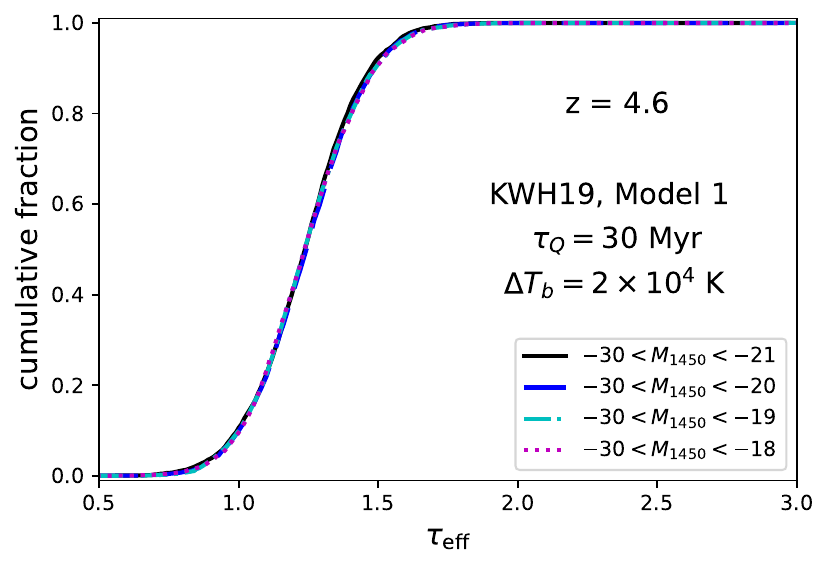}}
%\vspace{-1.5cm}
\caption{Distribution of effective optical depth averaged over regions $50h^{-1}$~cMpc across at $z=4.6$, shown for a UVB provided by the KWH19 Model 1 QSO luminosity function with the indicated absolute magnitude ranges.
   }
\label{fig:taueff_Mmax}
\end{figure}

While post-hydrogen reionization temperature fluctuations are disfavoured as an explanation for the broadenened optical depth distribution in cosmic reionization scenarios \citep{2021ApJ...923...87C}, the opposite may be expected for temperature fluctuations arising from QSO \HeIII\ regions, since a high opacity region with a low galaxy density may also have fewer QSOs and so cooler gas, allowing for a higher \HI\ fraction and so a higher opacity to \Lya\ photons. The higher opacity may also enhance the underdensity in detected Ly$\alpha$-emitting galaxies, as they may be significantly obscurred as a consequence of a locally higher \HI\ fraction resulting from a lower UVB flux from UVB fluctuations driven by QSOs \citep{2022MNRAS.516..572M}.

\begin{figure}
%\scalebox{0.5}{\includegraphics{fig01_a.eps}\includegraphics{fig01_b.eps}}
\scalebox{0.55}{\includegraphics{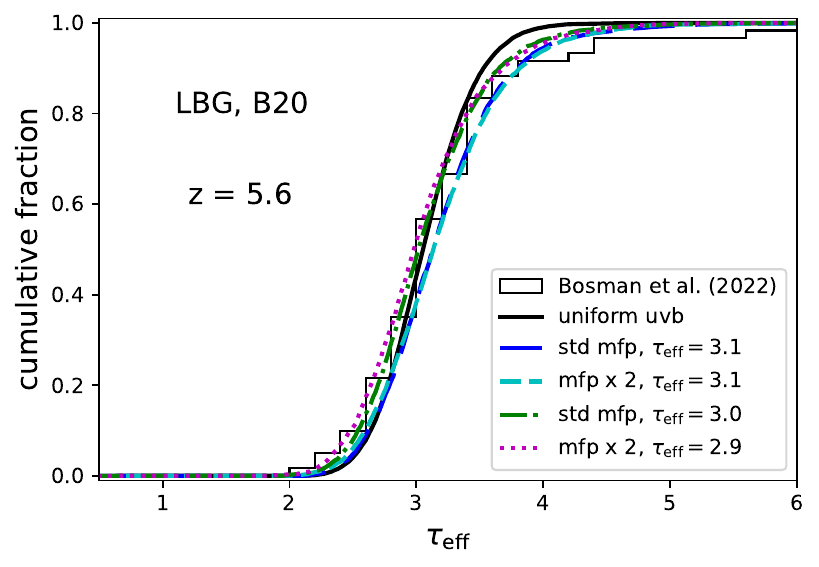}}
%\vspace{-1.5cm}
\caption{Distribution of effective optical depth averaged over regions $50h^{-1}$~cMpc across at $z=5.6$, shown for a UVB provided by the brightest LBGs with $f_\mathrm{esc}=1$ using the \citet{2021AJ....162...47B} luminosity function. Also shown is the case for a uniform UVB (black solid line). Measurements from \citet{2022MNRAS.514...55B} are shown for comparison. 
  }
\label{fig:taueff_LBGs}
\end{figure}

QSO sources dimmer than  $M_{1450}<-21$ are not detected by KWH19 at $z>1$, although this may be a result of incompleteness, as G19 detect AGN down to $M_{1450}<-19$ at redshifts as high as $z\sim5.6$ and the \emph{JWST} JADES survey to  $M_{1450}<-18$ at $z=4-6$ \citep{2023arXiv230801230M}. The escape fraction of ionizing radiation, especially radiation capable of ionizing \HeII, for such faint systems is unknown. None the less, motivated by the \emph{JWST} results, we explore the effects of extending the KWH19 Model 1 luminosity functions to fainter magnitudes, noting that the QSO density at $M_{1450}<-19$ for Model 1 over the redshift range $4.1 < z < 4.7$ approaches the QSO density measured in the JADES survey. The dimmer sources effectively substitute for galaxies, and so negligibly affect the \Lya\ optical depth distribution, as shown in Fig.~\ref{fig:taueff_Mmax}. (A $80h^{-1}$~cMpc simulation box is used.) A larger effect is found on the 1D flux power spectrum, as discussed below.

We have also considered the shot noise contribution to UVB fluctuations that may arise from galaxies. For the LBG luminosity function of \citet{2021AJ....162...47B}, an escape fraction of $\sim30$ percent of ionizing radiation may account for the UVB at $5<z<6$ to match the median \Lya\ optical depth (Fig.~\ref{fig:GamHI_LBG}). The resulting UVB is too uniform to much broaden the optical depth distribution, although possibly allowing for galaxy spatial correlations would further broaden the distribution. Including these is beyond the scope of this work. Instead, we consider the maximum shot-noise UVB fluctuations from the brightest LBGs consistent with a median optical depth $\tau_\mathrm{med}\simeq3.06$, which requires $M_\mathrm{AB, 1600}<-17.2$ for an escape fraction of unity. No diffuse recombination radiation from the IGM is included, even though Lyman Limit systems may account for as much as 40 percent of the total UVB \citep{MM93, 1996ApJ...461...20H}. Because no additional uniform UVB is assumed, we also take into account the hardening of the radiation from galaxies as filtered by the IGM to compute an accurate determination of the total UVB. Both of these work to maximise the fluctuations from the galaxies by minimising the number of sources needed to provide the UVB; this provides an upper limit to the UVB fluctuations galaxy shot noise may generate. Also considered is a mean free path for Lyman continuum photons twice the value of \citet{2021MNRAS.508.1853B} to allow for fewer sources, with $M_\mathrm{AB, 1600}<-18.7$, in an attempt to further enhance the UVB fluctuations. Slight variations in the magnitude cuts giving $\tau_\mathrm{med}\simeq3.0$ and 2.9 are also considered. The predicted optical depth distributions are broadened beyond the case of a uniform UVB, as shown in Fig.~\ref{fig:taueff_LBGs}, but none recover the distribution for both $\tau_\mathrm{eff}<2.5$ and $\tau_\mathrm{eff}>4$. Such a scenario is in any case disfavoured by the anti-correlation between escape fraction and galaxy luminosity inferred from the spectral modelling of observed galaxies \citep{2022MNRAS.517.5104C}. None the less, the contribution from bright LBGs, if their escape fractions are high, may further enhance the breadth of the distribution in conjunction with QSO sources.

\subsection{Pixel flux auto-correlation function and power spectrum}
\label{subsec:xiPkF}

It has long been recognised that QSOs may provide a large shot-noise contribution to spatial correlations in the UVB \citep{1992MNRAS.258...45Z}. These correlations in turn enhance spatial correlations in the \HI\ fraction on large scales, which should appear as enhanced correlations on large scales in the pixel fluxes measured in the \Lya\ forest \citep{2004ApJ...610..642C, 2004MNRAS.350.1107M, 2014MNRAS.442..187G, 2014PhRvD..89h3010P, 2019MNRAS.482.4777M}. 

Measurements of the \Lya\ flux auto-correlation function show values near unity at $5<z<6$ for separations of up to several comoving megaparsecs \citep{2024MNRAS.531.3069W}. These are not well reproduced by simulations with a uniform UVB for separations beyond about $1.5h^{-1}$~cMpc, although the measurement errors are moderately large. We find similarly weaker correlations allowing for QSOs following the KWH19 luminosity functions. By contrast, drawing QSOs from the G19 luminosity function well recovers the strong correlations for the same \HeIII-region temperature boosts that well reproduce the width of the \Lya\ optical depth distribution at $z\simeq5.6$.

\begin{figure*}
%\scalebox{0.5}{\includegraphics{fig01_a.eps}\includegraphics{fig01_b.eps}}
\scalebox{0.6}{\includegraphics{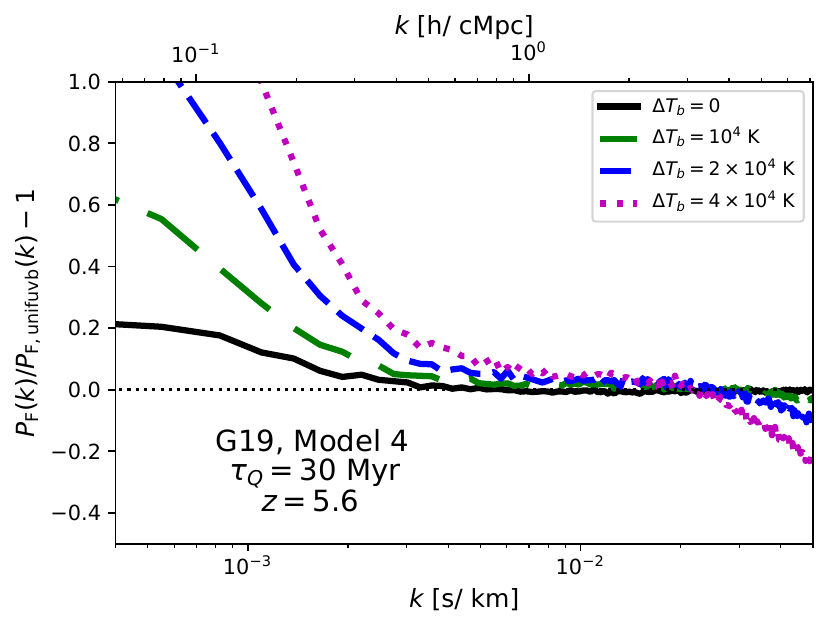}\includegraphics{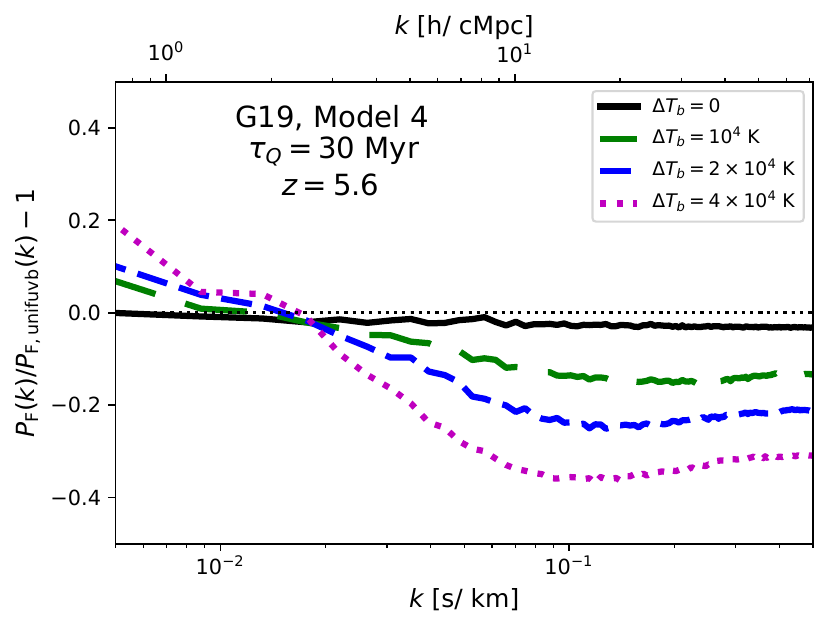}}
%\vspace{-1.5cm}
\caption{The relative difference in the pixel flux power spectrum at $z=5.6$ including QSOs following the G19 Model 4 QSO luminosity function, for the indicated \HeIII-region temperature boosts $\Delta T_b$ and with a QSO lifetime $\tau_Q=30$~Myr, compared with a homogeneous UVB. {\it Left-hand panel}:\ $160h^{-1}$~cMpc box. {\it Right-hand panel}:\ $10h^{-1}$~cMpc box. 
}
\label{fig:dPkF_dTb_G19}
\end{figure*}

\begin{figure}
%\scalebox{0.5}{\includegraphics{fig01_a.eps}\includegraphics{fig01_b.eps}}
\scalebox{0.55}{\includegraphics{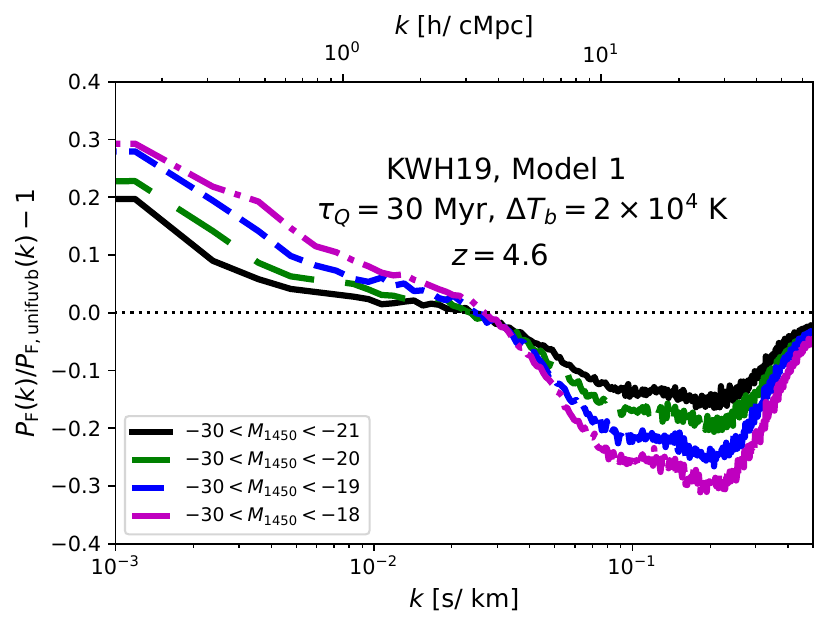}}
%\vspace{-1.5cm}
\caption{The relative difference in the pixel flux power spectrum at $z=4.6$ including QSOs following the KWH19 Model 1 QSO luminosity function for a \HeIII-region temperature boost $\Delta T_b=2\times10^4$~K and the indicated faint QSO magnitude limits, compared with a homogeneous UVB. A $40h^{-1}$~cMpc simulation box is used.
}
\label{fig:dPkF_Mmax_KWH19}
\end{figure}

Similarly, the shot noise contribution to the UVB will enhance the pixel flux power spectrum at low wavenumbers ($k\lsim0.02\,\mathrm{s\,km^{-1}}$). While the enhancement was typically found at the few percent level at $z\sim3$ in simulations without inhomogeneous temperatures \citep{2004ApJ...610..642C, 2004MNRAS.350.1107M}, allowing for the temperature fluctuations associated with QSO \HeIII\ regions much magnifies the enhancement at $z\sim4.6$, depending on both the temperature boost in the \HeIII\ regions and their intergalactic filling factor, even for the KWH19 QSO luminosity functions. Allowing for the G19 luminosity function at $z\sim5.6$ results in still larger enhancements, both with and without the associated temperature fluctuations, as shown in the left-hand panel of Fig.~\ref{fig:dPkF_dTb_G19}. Extending the KWH19 Model 1 QSO luminosity function to fainter magnitudes $M_{1450}>-21$ similarly enhances the large-scale power at $z=4.6$, as shown in Fig.~\ref{fig:dPkF_Mmax_KWH19}, while increasing the suppression of power on small scales. The strengths of the effects increase with the \HeIII\ porosity parameter, which increases as increasingly fainter QSOs are included (Fig.~\ref{fig:QHeIII_Mmax}).

At higher wavenumbers ($k\gsim0.02\,\mathrm{s\,km^{-1}}$), UVB fluctuations alone have a negligible effect on the flux power spectrum. Allowing for the induced temperature inhomogeneities arising from the \HeIII\ regions suppresses the power by tens of percent, with the magnitude of suppression again increasing both with the temperature boost and the \HeIII-region filling factor. We considered representative soft ($\alpha_Q=1.57$) and hard ($\alpha_Q=0.61$) QSO spectral indices, as well as a combination of both with faint sources harder than bright. Alternative spectral indices primarily affect the results through the change in the \HeIII\ porosity parameter. For example, choosing $\alpha_Q=1.96$ \citep{1997ApJ...475..469Z} for the G19 Model 4 QSO luminosity function case reduces the \HeIII\ porosity from 0.22 for $\alpha_Q=1.57$ to 0.11 at $z=5.6$, for a source lifetime $\tau_Q=30$~Myr. The resulting \Lya\ optical depth distribution for a temperature boost $\Delta T_b=2\times10^4$~K is only slightly broader than the distribution for $\alpha_Q=1.57$ and no temperature boost, while the 1D \Lya\ flux power spectrum suppression at $k=0.1\,\mathrm{s\,km^{-1}}$ is reduced (in magnitude) from 21 percent to 13 percent.

The overall effect on the flux power spectrum is strikingly similar to the effect of inhomogeneous reionization as found in the Sherwood-Relics simulations \citep{2023MNRAS.519.6162P}. This is not surprising, in that in both cases, small scale suppression in power will result from broadened absorption lines in heated regions, while the large scale enhancement in power is a consequence of large-scale inhomogeneities in the IGM temperature. A difference is that in the Sherwood-Relics simulations, the transition between large-scale enhancement in power and small-scale suppression occurs at about twice the wavenumber ($k\simeq0.04\,\mathrm{s\,km^{-1}}$), found here from \HeIII\ reionization. This may, however, depend on the reionization scenario \citep{2019MNRAS.490.3177W, 2019MNRAS.486.4075O}. Hydrodynamical feedback and the associated pressure-smoothing are also not included here. None the less, our findings suggest the UVB and temperature fluctuations associated with a high number of QSOs at $5<z<6$ as suggested by the G19 luminosity function present an alternative to a late completion scenario for reionization at $z\simeq5.3$. More generally, surviving temperature inhomogeneities from hydrogen reionization and large numbers of QSOs may work in conjunction to broaden the \Lya\ optical depth distribution at $z>5.5$ even if the reionization of hydrogen completed at $z\sim6$.

Other statistics may offer additional discriminatory tests on the role of QSOs at $z>5$, such as the distribution of \Lyb\ optical depths \citep{2020MNRAS.497..906K},  \Lya\ and \Lyb\ transmission spikes in the spectra of the \Lya\ and \Lyb\ forests \citep{2017A&A...601A..16B, 2018MNRAS.473..765C, 2018MNRAS.479...43K, 2020MNRAS.494.5091G, 2020ApJ...904...26Y} and damping wings in the spectra of high redshift QSOs \citep{2024MNRAS.533.1525B}. The temperature distribution of the IGM would also indicate recent reionization, either ongoing hydrogen reionization or early \HeII\ reionization. Estimates using transmission spikes suggest a temperature of $8000\lsim T\lsim 13000$~K at $5.4<z<5.8$ \citep{2020MNRAS.494.5091G}, while \Lya\ forest flux power spectra measurements suggest $5000\lsim T\lsim 7000$~K at $5<z<5.4$ \citep{2019ApJ...872...13W}, smaller than may be expected from either early QSO reionization of \HeII\ or a late end to hydrogen reionization. Such lower temperatures are more consistent with hydrogen reionization completing by $z\simeq6$ rather than $z\simeq5.3$ \citep{2022ApJ...933...59V}. It must be recognised, however, that there are no direct temperature measurements of the diffuse IGM, only proxies calibrated through numerical simulations. Allowing for UVB and thermal fluctuations or a late completion time to reionization would require new calibrating simulations, or direct comparison of these simulations with observations.
  
\section{Conclusions}
\label{sec:conc}

Once QSO sources form in abundance at high redshifts, they will influence the \HI\ component of the IGM both through the \HI\ photoionization rate and a boosted IGM temperature within \HeIII\ regions. Estimates of the influence on the \HI\ \Lya\ forest are hampered by the uncertainties in the luminosity function, spectra and lifetimes of QSOs. Assessing the magnitudes of the effects is computationally challenging, as it involves radiative transfer of ionizing radiation over length scales of several megaparsecs. To estimate the influence on 1-point and 2-point statistics of the \HI\ \Lya\ forest, we use instead a semi-analytic approach for the growth of \HeIII\ regions, noting more precise assessments must await simulations with full radiative transfer. We summarise the main conclusions here.

\begin{itemize}
\item The growth rate of the filling factor of \HeIII\ is sensitive to the lifetime of the sources (Fig.~\ref{fig:QHeIII}). This arises from the restriction of the growth rate of the ionization front by causality, as QSO sources are so bright all the emitted photons are not immediately absorbed by the surrounding gas in the early expansion phase. As a consequence, the \HeIII\ ionization fronts of longer-lived QSOs take longer to catch up to the radius within which all emitted photons have been absorbed.

\item For the \citet{2019MNRAS.488.1035K} (KWH19) QSO luminosity function, based on optical-infrared selected QSOs, the \HeIII\ regions induce weak spatial correlations in the IGM temperature field at high redshifts ($z\simeq5.6$), declining to under a few percent for lag distances beyond a few comoving Mpc, but may persist at over 1 percent for lag distances as great as $10h^{-1}$~cMpc (Fig.~\ref{fig:xiT}, left-hand panel).

\item For the \citet{2019ApJ...884...19G} (G19) QSO luminosity function, based on x-ray selected QSOs, the \HeIII\ regions induce spatial correlations in the IGM temperature field at high redshifts ($z\simeq5.6$), that may persist at over 10 percent for lag distances as great as $10h^{-1}$~cMpc (Fig.~\ref{fig:xiT}, right-hand panel).

\item QSOs induce spatial correlations in the metagalactic photoionizing background (including galaxies) with a correlation length of a few comoving Mpc at high redshifts (Fig.~\ref{fig:xiG}), persisting at a level of 1 percent for lag separations as great as $20h^{-1}$~Mpc, with approximately equal strength for both the KWH19 and G19 QSO luminosity functions.

\item QSOs following the KWH19 QSO luminosity functions negligibly broaden the \HI\ optical depth distribution smoothed over $50h^{-1}$~cMpc intervals at $z\simeq5.6$ compared with a uniform UVB (Fig.~\ref{fig:taueff_KWH19}), and fail to match the observed distribution.

\item QSOs following the G19 QSO luminosity functions significantly broaden the \HI\ optical depth distribution smoothed over $50h^{-1}$~cMpc intervals at $z\simeq5.6$ compared with a uniform UVB  (Fig.~\ref{fig:taueff_G19}), with the width of the distribution increasing with the \HeIII\ temperature boost, and may even over-produce the width for \HeIII\ temperature boosts as high as $4\times10^4$~K, depending on the QSO lifetime. For a given \HeIII\ temperature boost, the width of the distribution narrows for long QSO lifetimes ($\tau_Q\ge300$~Myr) because of the reduced \HeIII\ filling factor and associated IGM temperature fluctuations. While QSOs may readily reproduce the broadened distribution for low effective \HI\ optical depths $\tau_\mathrm{eff}$, models do not readily recover values as high as $\tau_\mathrm{eff}>4.5$ without overly broadening the distribution for low $\tau_\mathrm{eff}$. Allowing for hard QSO spectra shortward of the Lyman edge for dim QSOs, while soft for bright QSOs, however, matches the distribution at $z\simeq5.6$ for both low and high values of $\tau_\mathrm{eff}<5.0$ (Fig.~\ref{fig:taueff_G19hy}). The models none the less still fall short of the observed fraction of lines of sight with $\tau_\mathrm{eff}>5.0$.

\item Allowing for an escape fraction of unity from the brightest Lyman Break Galaxies broadens the \HI\ optical depth distribution smoothed over $50h^{-1}$~cMpc intervals at $z\simeq5.6$ compared with a uniform UVB, but not sufficiently to match the breadth of the measured distribution (Fig.~\ref{fig:taueff_LBGs}). Such an extreme model may none the less contribute to the breadth of the distribution along with QSO sources.

\item The \HI\ \Lya\ pixel flux auto-correlation function allowing for QSOs following the KWH19 QSO luminosity functions (Fig.~\ref{fig:xiF}, left-hand panel), matches the measured pixel flux auto-correlation function at $z=5.6$ only on scales smaller than $\sim1.5h^{-1}$~cMpc, and declines too quickly for larger separations compared with the mean measured values regardless of the size of the \HeIII-region temperature boost.

\item The pixel flux auto-correlation function allowing for QSOs following the G19 QSO luminosity functions (Fig.~\ref{fig:xiF}, right-hand panel), matches the measured pixel flux auto-correlation function at $z=5.6$ on scales smaller than $\sim1.5h^{-1}$~cMpc regardless of the size of the \HeIII-region temperature boost. The predictions at larger separations for temperature boosts of $1-2\times10^4$~K recover the mean measured values.

\item We confirm that allowing for UVB fluctuations from QSO sources, without a temperature boost in \HeIII\ regions, enhances the flux power spectrum on large scales ($k<0.2h\,\mathrm{cMpc}^{-1}$) by a few percent at $4<z<5$ (Fig.~\ref{fig:dPkF_dTb}).

\item Allowing for both UVB fluctuations and boosts in the \HeIII\ region temperatures enhances the flux power spectrum on scales $k<0.02\,\mathrm{s\,km^{-1}}$ ($k\lsim2h\,\mathrm{cMpc}^{-1}$) at $z>4$ (Figs.~\ref{fig:dPkF_dTb} and \ref{fig:dPkF_dTb_G19}, left-hand panels). For $k<0.001\,\mathrm{s\,km^{-1}}$ ($k\lsim0.1h\,\mathrm{cMpc}^{-1}$), the enhancement is by as much as 50 percent. On scales $0.02<k\lsim0.5\,\mathrm{s\,km^{-1}}$, power is suppressed by as much as 30 percent (Figs.~\ref{fig:dPkF_dTb} and \ref{fig:dPkF_dTb_G19}, right-hand panels). The magnitude of both effects increases both with the \HeIII-region temperature boost and the filling factor of the \HeIII\ regions in the IGM, and so dependent on QSO lifetime (Fig.~\ref{fig:dPkF_models}), cosmic time (Fig.~\ref{fig:dPkF_evol}) and the minimum luminosity of QSOs able to photoionize \HeII\ (Fig.~\ref{fig:dPkF_Mmax_KWH19}).

\end{itemize}

The success of the G19 QSO luminosity function for recovering both the \Lya\ optical depth distributions at $z>5$ and the pixel flux auto-correlation function suggest QSO-induced UVB and temperature fluctuations provide an alternative scenerio to the late completion of reionization at $z\simeq5.3$, although regions with $\tau_\mathrm{eff}>5$ may indicate residual neutral regions persist to late times. Additional tests may be more discriminatory, such as the \Lyb\ optical depth distribution, the distribution of peaks in \Lya\ forest spectra and damping wing signatures in the spectra of QSOs.

The modelling could be improved by taking into account the variable mean free path to Lyman continuum photons that must accompany UVB fluctuations, the evolution of \HeIII\ regions, including their temperature and hydrodynamical response, and a range in QSO spectral indices and lifetimes. Spatial correlations in the sources would also contribute to the fluctuations in the UVB and in the IGM temperature and \HI\ fraction. The impact of QSO sources may also be incorporated into inhomogeneous hydrogen reionization models. Fully self-consistent modelling would require radiative transfer simulations, which remain computationally expensive. Until such complete computations become feasible, Monte Carlo approaches with semi-analytic radiative transfer as here may help to provide some insight into how uncertainties in the properties and numbers of QSOs affect estimates of the impact of QSOs on the IGM.

\section*{Acknowledgments}
The authors thank James Bolton for comments on an earlier draft. AM thanks Laura Keating for discussions and assistance in accessing the Sherwood-Relics simulation data, and Stefano Cristiani and Matteo Viel for discussions. AM thanks SISSA and the INAF-Astronomical Observatory of Trieste for their hospitality where much of this work was carried out. Both authors thank the referee for helpful comments that improved the paper. The Sherwood-Relics simulations were performed using the Joliot Curie supercomputer at Le Tr\`es Grand Centre de Calcul (TGCC), the DiRAC Data Intensive service (CSD3) at the University of Cambridge, and the DiRAC Memory Intensive service Cosma6 at Durham University. CSD3 is managed by the University of Cambridge University Information Services on behalf of the STFC DiRAC HPC Facility (www.dirac.ac.uk). The DiRAC component of CSD3 at Cambridge was funded by BEIS, UKRI and STFC capital funding and STFC operations grants. Cosma6 is managed by the Institute for Computational Cosmology on behalf of the STFC DiRAC HPC Facility (www.dirac.ac.uk). The DiRAC service at Durham was funded by BEIS, UKRI and STFC capital funding, Durham University and STFC operations grants. DiRAC is part of the UKRI Digital Research Infrastructure. 

\section*{Data Availability}
No new observational data were generated or analysed in support of this research.

\section*{Rights Retention Statement}

For the purpose of open access, the author has applied a Creative Commons Attribution (CC BY) licence to any Author Accepted Manuscript version arising from this submission.

\bibliographystyle{mn2e-eprint}
\bibliography{ms}

\appendix 

\section{Convergence tests}
\label{ap:conv}

\subsection{Convergence tests of the effective \Lya\ optical depth distribution}
\label{ap:taueff_conv}

Convergence tests of the effective optical depth distribution are presented here for the G19 QSO luminosity function with \HeIII\ region temperature boost $\Delta T_b=2\times10^4$~K. 

\begin{figure}
%\scalebox{0.5}{\includegraphics{fig01_a.eps}\includegraphics{fig01_b.eps}}
\scalebox{0.55}{\includegraphics{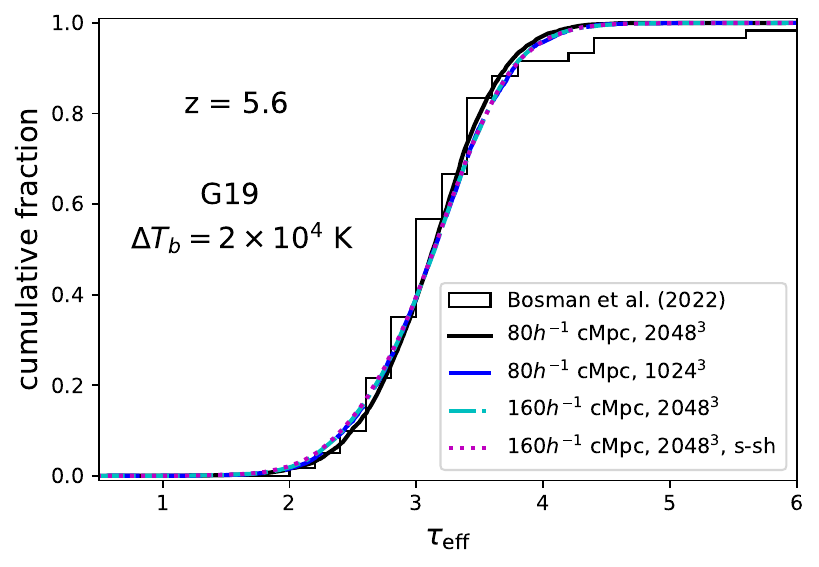}}
%\vspace{-1.5cm}
  \caption{Convergence with box size and spatial resolution of the distribution of effective \Lya\ optical depth averaged over regions $50h^{-1}$~cMpc across. A QSO lifetime $\tau_Q=100$~Myr is assumed. Shown for the G19 Model 4 luminosity function for a temperature boost $\Delta T_b=2\times10^4$~K, and a QSO spectral index $\alpha_Q=1.57$. Results for box sizes of $80h^{-1}$ and $160h^{-1}$~cMpc are shown, with initial gas particle numbers of   $1024^3$ or $2048^3$, as indicated. Also shown is a case including self-shielding (\lq s-sh\rq; see text). Measurements from \citet{2022MNRAS.514...55B} are shown for comparison. Increasing the box size broadens the distribution. 
  }
\label{fig:taueff_conv_G19}
\end{figure}  
The convergence of the effective optical depth distribution at $z=5.6$, for optical depths averaged over intervals $50h^{-1}$~cMpc across, is shown in Fig.~\ref{fig:taueff_conv_G19} for box sizes of $80h^{-1}$ and $160h^{-1}$~cMpc, and gas particle numbers of $1024^3$ and $2048^3$. A QSO lifetime $\tau_Q=100$~Myr is assumed. While the results are well converged with the number of gas particles, increasing the box size at fixed mean mass resolution slightly broadens the distribution. Since dense regions may become self-shielded from the UVB, and so increase the \Lya\ optical depth, a case is also shown (magenta dotted line) allowing for self-shielding using the approximate \HI\ fraction correction method of \citet{2013MNRAS.430.2427R}. Self-shielding has a negligible effect on the optical depth distribution.

\begin{figure}
%\scalebox{0.5}{\includegraphics{fig01_a.eps}\includegraphics{fig01_b.eps}}
\scalebox{0.55}{\includegraphics{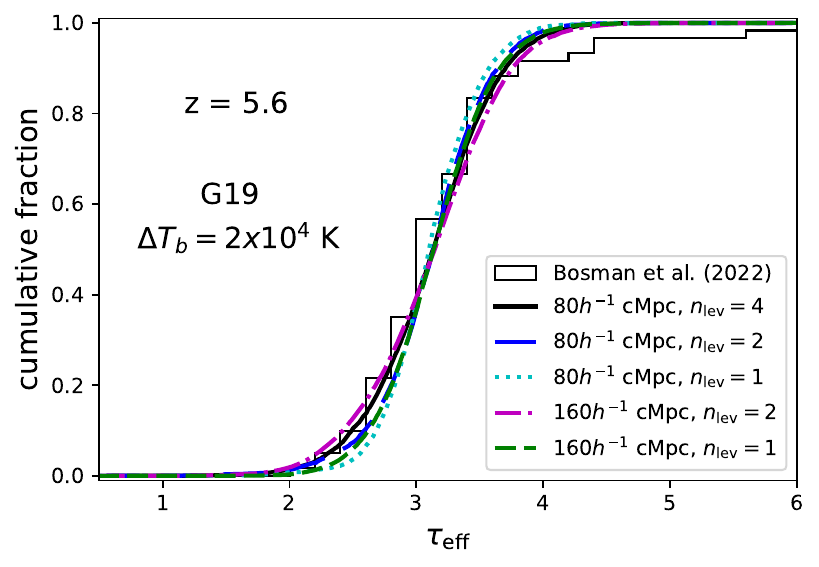}}
%\vspace{-1.5cm}
\caption{Convergence of the distribution of effective \Lya\ optical depth averaged over regions $50h^{-1}$~cMpc across with the number of levels of surrounding box extensions. Shown for the G19 Model 4 luminosity function for a temperature boost $\Delta T_b=2\times10^4$~K, QSO lifetime $\tau_Q=100$~Myr and a QSO spectral index $\alpha_Q=1.57$. Results for box sizes of $80h^{-1}$ and $160h^{-1}$~cMpc are shown, with the indicated levels of box extensions. Measurements from \citet{2022MNRAS.514...55B} are shown for comparison. The distribution broadens with an increasing number of box levels.
  }
\label{fig:taueff_nboxconv_G19}
\end{figure}
The convergence of the effective optical depth distribution at $z=5.6$ is shown in Fig.~\ref{fig:taueff_nboxconv_G19} for simulation box sizes of $80h^{-1}$ and $160h^{-1}$~cMpc, both with $2048^3$ gas particles, for varying levels of surrounding boxes for QSOs, as indicated. The full extension corresponds to $n_\mathrm{lev}=4$ for the $80h^{-1}$~cMpc simulation (with full box side length $560h^{-1}$~cMpc) and $n_\mathrm{lev}=2$ for the $160h^{-1}$~cMpc simulation (with full box side length $480h^{-1}$~cMpc), covering the redshift range $z=5.6$ for the original simulation volume to $z\sim7$ for the largest surrounding box within the causal horizon of the simulation volume. The tests show that good convergence, particularly at low $\tau_\mathrm{eff}$ values, requires the full extension of surrounding boxes.

\begin{figure}
%\scalebox{0.5}{\includegraphics{fig01_a.eps}\includegraphics{fig01_b.eps}}
\scalebox{0.55}{\includegraphics{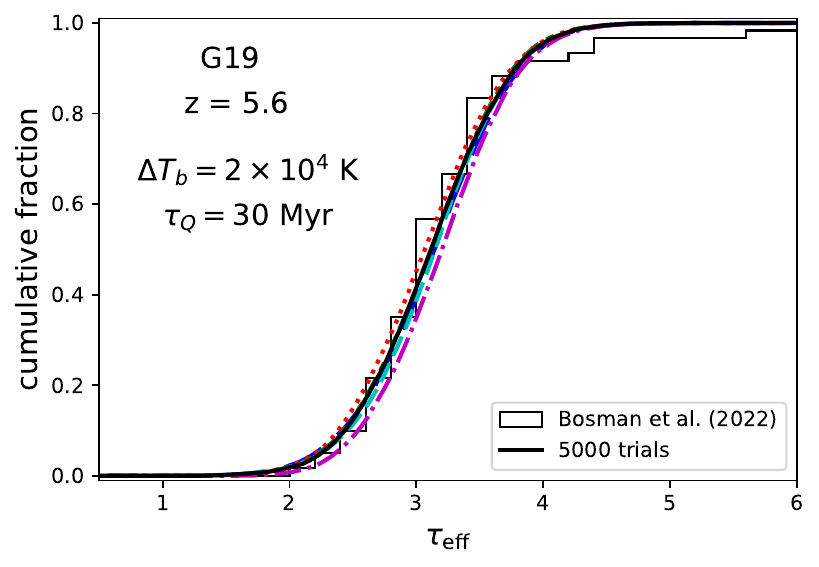}}
%\vspace{-1.5cm}
\caption{The distribution of effective \Lya\ optical depth averaged over regions $50h^{-1}$~cMpc across, and as averaged over 5000 lines of sight through a $160h^{-1}$~cMpc box. Shown for the G19 Model 4 luminosity function for a temperature boost $\Delta T_b=2\times10^4$~K, QSO lifetime $\tau_Q=30$~Myr and a QSO spectral index $\alpha_Q=1.57$. Results are shown when taking an independent QSO population random realisation per line of sight (black solid line), and for five separate QSO population realisations (coloured lines), each averaging over the same 5000 lines of sight. Measurements from \citet{2022MNRAS.514...55B} are shown for comparison.
}
\label{fig:taueff_Nran1_G19}
\end{figure}
Throughout this paper, the optical depths are averaged using 5000 fixed lines of sight through a simulation volume. To ensure statistical independence of the radiation field for each line of sight, a separate random realisation of the QSO population is performed for each line of sight. Fig.~\ref{fig:taueff_Nran1_G19} compares the average distribution at $z=5.6$ for QSOs drawn from the G19 Model 4 QSO luminosity function using this method (black solid line) against sample distributions each derived from a separate single realisation of the QSO population used for averaging the optical depth distribution over all 5000 lines of sight; five such cases are illustrated (coloured lines), all using the same uniform contribution to the total UVB, in addition to the QSOs. While the distributions are similar, a moderate amount of spread results. The main effect of using a single random realisation of the QSO population may be a re-calibration of the magnitude of the hydrogen photoionization rate $\Gamma_\mathrm{HI}$ required to match the median effective optical depth $\tau_\mathrm{med}$. We find the required photoionization rate scales approximately as $\Gamma_\mathrm{HI}\sim\tau_\mathrm{med}^{1.9-2.5}$, so that a typical single random realisation of the QSO population may result in a mis-estimate of $\Gamma_\mathrm{HI}$ by 20--30 percent.

\begin{figure}
%\scalebox{0.5}{\includegraphics{fig01_a.eps}\includegraphics{fig01_b.eps}}
\scalebox{0.55}{\includegraphics{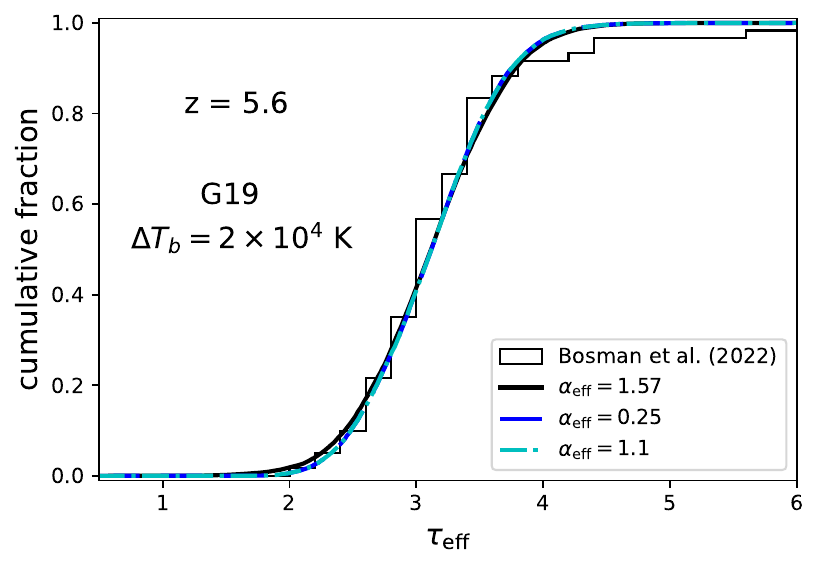}}
%\vspace{-1.5cm}
\caption{The distribution of effective \Lya\ optical depth averaged over regions $50h^{-1}$~cMpc across at $z=5.6$ for the indicated effective spectral indices shortward of the hydrogen Lyman edge for computing the photoionization rate. Shown for the G19 Model 4 QSO luminosity function for a temperature boost $\Delta T_b=2\times10^4$~K and QSO lifetime $\tau_Q=30$~Myr. Measurements from \citet{2022MNRAS.514...55B} are shown for comparison. 
  }
\label{fig:taueff_G19_aQLL}
\end{figure} 
Allowing for filtering of the intrinsic spectrum of the QSO by photoelectric absorption in the IGM negligibly affects the effective \Lya\ optical depth distribution when normalised to the same median value, as shown in Fig.~\ref{fig:taueff_G19_aQLL}. The alternative spectral indices $\alpha_\mathrm{eff}=0.25$ and 1.1 correspond to filtering of the intrinsic QSO spectral index $\alpha_Q=1.57$ by a diffuse IGM or \HI\ absorption systems with column density exponent $\beta=1.2$, respectively, as discussed in Sec.~\ref{subsec:mc}. 

\begin{figure}
%\scalebox{0.5}{\includegraphics{fig01_a.eps}\includegraphics{fig01_b.eps}}
\scalebox{0.55}{\includegraphics{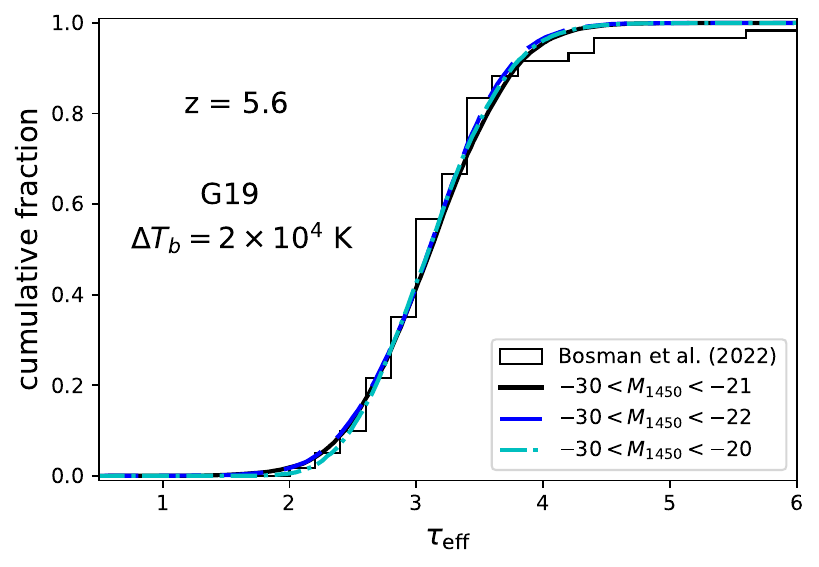}}
%\vspace{-1.5cm}
\caption{The distribution of effective \Lya\ optical depth averaged over regions $50h^{-1}$~cMpc across at $z=5.6$ for the indicated maximum QSO magnitudes for the sampled luminosity function. Shown for the G19 Model 4 QSO luminosity function for a temperature boost $\Delta T_b=2\times10^4$~K, QSO lifetime $\tau_Q=30$~Myr and a QSO spectral index $\alpha_Q=1.57$. Measurements from \citet{2022MNRAS.514...55B} are shown for comparison. 
  }
\label{fig:taueff_G19_Mmax}
\end{figure} 
Allowing for alternative maximum QSO magnitudes for the sampled QSO luminosity function has little effect on the effective \Lya\ optical depth distribution when normalised to the same median value, as shown in Fig.~\ref{fig:taueff_G19_Mmax}. A slight narrowing of the distribution for low $\tau_\mathrm{eff}$ values results for a maximum magnitude $M_{1450}<-20$ because of the near doubling in the the number of low luminosity QSOs compared with $M_{1450}<-21$, and so a slightly diminished range in the fluctuations of the UVB.

\subsection{Convergence tests of the UVB photoionization rate auto-correlation function}
\label{ap:xiG_conv}

Convergence tests of the auto-correlation function of the photoionization rate are presented here for the KWH19 and G19 QSO luminosity functions at $z=5.6$, with \HeIII\ region temperature boost $\Delta T_b=10^4$~K. A QSO lifetime $\tau_Q=30$~Myr is assumed.

\begin{figure*}
%\scalebox{0.5}{\includegraphics{xiG_KWH19_conv_z5.6.pdf}\includegraphics{xiG_G19_c onv_z5.6.pdf}}
\scalebox{0.6}{\includegraphics{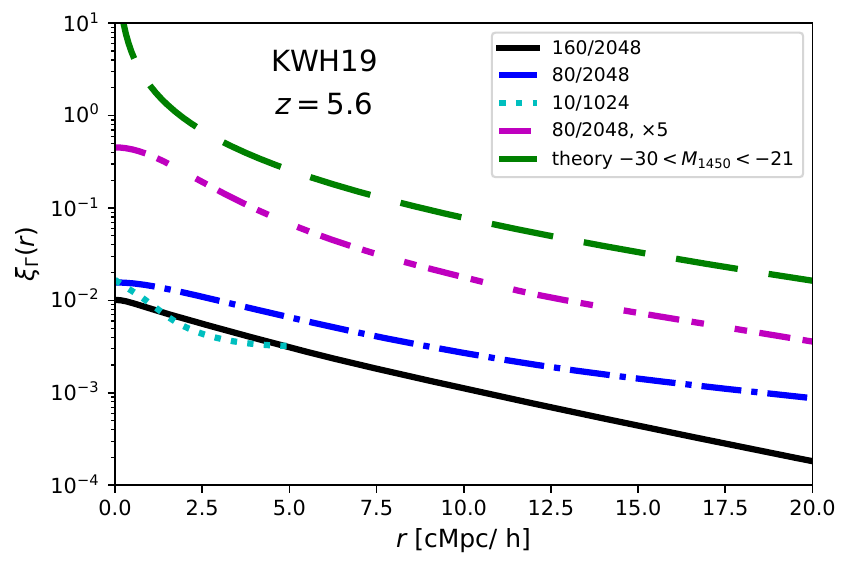}\includegraphics{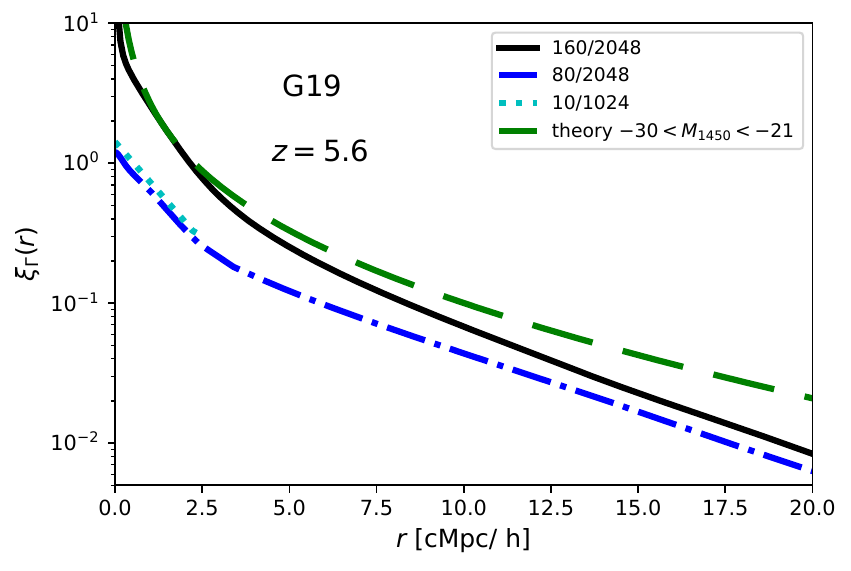}}
%\scalebox{0.5}{\includegraphics{xiG_KWH19_conv_z5.6.eps}\includegraphics{xiG_G19_conv_z5.6.eps}}
%\scalebox{0.5}{\includegraphics{xiG_KWH19_conv_z5.6.pdf}}
%\vspace{-1.5cm}
\caption{Convergence with box size of the total photoionization rate auto-correlation function at $z=5.6$. A QSO spectral index $\alpha_Q=1.57$, source lifetime $\tau_Q=30$~Myr and  temperature boost $\Delta T_b=10^4$~K are assumed in both panels. {\it Left-hand panel}:\ Results for the KWH19 Model 1 QSO luminosity function. {\it Right-hand panel}:\ Results for the G19 Model 4 QSO luminosity function. For both luminosity functions, QSOs are drawn from the absolute magnitude range $-30<M_{1450}<-21$ over 5000 Monte Carlo realisations each. Results are for simulation box sizes of 10, 80 and $160h^{-1}$~cMpc, as indicated. Theoretical predictions following \citet{1992MNRAS.258...45Z} are shown for comparison. Also shown in the left-hand panel is the result for the $80h^{-1}$~Mpc box with the numbers of QSOs from the KWH19 luminosity function boosted by a factor of 5 (magenta dot-dot-dashed line).
  }
\label{fig:xiG_KWH19_G19_conv}
\end{figure*}
The convergence of the total photoionization rate auto-correlation function, including contributions from both a uniform UVB and QSOs, at $z=5.6$ is shown in Fig.~\ref{fig:xiG_KWH19_G19_conv} for box sizes of 10, 80 and $160h^{-1}$~cMpc. The full level of encompassing QSO boxes along the backward light cone is used (see Sec.~\ref{subsec:mc}). For comparison, also shown are the infinite-volume theoretical predictions for constant comoving QSO luminosity functions following \citet{1992MNRAS.258...45Z}.
The auto-correlation function formally diverges at zero-lag for point sources.

The results for the KWH19 QSO luminosity function (Model 1) (including the uniform UVB contribution) are fairly well converged, but at a much lower level than the theoretical prediction, as shown in the left-hand panel. By contrast, the auto-correlation function for the G19 QSO luminosity function is fairly well converged to the expected strength for the largest box simulation, especially at small separations, as shown in the right-hand panel. At large separations, $r>15h^{-1}$~cMpc, the result falls short of the theoretical expectation, consistent with the slow convergence at large separations generally found in simulations for spatial auto-correlation functions \citep{2004MNRAS.350.1107M}.

The reason for the anomalously low strength of the correlations for the KWH19 luminosity function is not fully clear. Since the correlations arise from source shot noise, the amplitude of the correlations scales inversely with the effective number density of the sources $n_\mathrm{eff}$, defined by $n_\mathrm{eff}=\langle L\rangle^2/\langle L^2\rangle$ \citep{1992MNRAS.258...45Z}, where the averages are weighted by the source luminosity density $\Phi(L)$:\ $\langle(\dots)\rangle=\int (\dots)\Phi(L)\,dL$. The result from the $160h^{-1}$~cMpc simulation is $1/n_\mathrm{eff}=1.1\times10^4$~cMpc$^3$, in good agreement with the theoretical expectation for the KWH19 luminosity function of $1.0\times10^4$~cMpc$^3$. The luminosity variance $\langle L^2\rangle$ is dominated by the QSOs \citep{2020MNRAS.491.4884M}. For the KWH19 luminosity function, 90 percent of the correlation strength (including the galactic contribution) is carried by QSOs with $M_{1450}<-26.4$ (i.e., limiting the QSO magnitude range to $-26.4<M_{1450}<-21$ reduces the expected correlation strength by a factor of 10). A comoving box size of 1100~cMpc is required to contain a single QSO with $M_{14.50}<-26.4$. This suggests there may be substantial cosmic variance in the UVB correlation function. Moreover, the evolution in the QSO luminosity function would need to be included, which the formulation of \citet{1992MNRAS.258...45Z} does not account for.

The sensitivity of the correlations to QSO numbers is confirmed by artificially boosting the numbers of QSOs by a factor of 5. The correlations should strengthen by a factor of 5 (since the mean UVB is dominated by the uniform component, so $1/n_\mathrm{eff}$ increases by a factor of $\sim5$), while the correlations are found to strengthen by a factor of $\sim20$, as shown in Fig.~\ref{fig:xiG_KWH19_G19_conv}. In many random realisations, the bright QSOs are simply missing and so do not contribute to the average strength of the correlations. By comparison, using instead the G19 luminosity function, 90 percent of the correlation strength (including galaxies) is carried by QSOs with $M_{1450}<-24.9$, for which a comoving box size of only 260~cMpc is required to include a single QSO, so that there are many QSOs within the horizon to contribute to the shot noise component of the UVB fluctuations. The correlation strength from the $160h^{-1}$~cMpc box simulation agrees well with the theoretical expectation (right-hand panel of Fig.~\ref{fig:xiG_KWH19_G19_conv}). The weaker results for the smaller boxes may be a consequence of the shorter lines of sight for estimating the correlations, as they less well sample the UVB correlations. We leave the possibility of large cosmic variance in the UVB correlation function when the QSO numbers are extremely low as an open question.

\subsection{Convergence tests of the \Lya\ pixel flux power spectrum}
\label{ap1:PkF_conv}

\begin{figure}
%\scalebox{0.5}{\includegraphics{fig01_a.eps}\includegraphics{fig01_b.eps}}
%\scalebox{0.5}{\includegraphics{dFluxPk_iQmod1_tauQ30Myr_conv_z4.6.eps}}
%\scalebox{0.5}{\includegraphics{dFluxPk_iQmod1_tauQ30Myr_conv_z4.6.pdf}}
\scalebox{0.55}{\includegraphics{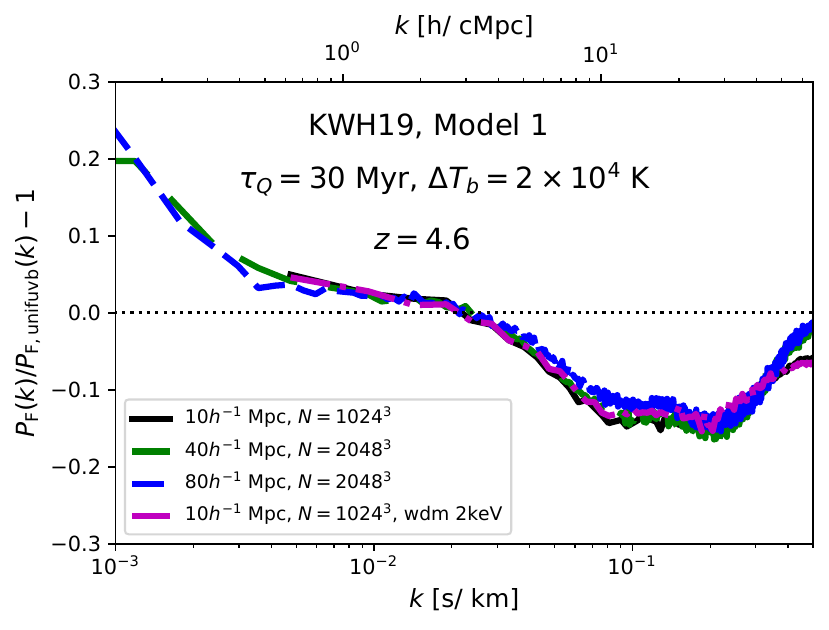}}
%\vspace{-1.5cm}
\caption{The relative change in the \Lya\ pixel flux power spectrum for the KWH19 QSO Model 1 luminosity function at $z=4.6$ for a \HeIII-region temperature boost $\Delta T_b=2\times10^4$~K for the indicated box sizes and mass resolutions.
}
\label{fig:PkF_conv}
\end{figure} 

Fig.~\ref{fig:PkF_conv} shows the relative change in the \Lya\ flux power spectrum $\delta P_F(k)$ at $z=4.6$ allowing for QSO \HeIII-regions with a temperature boost of $\Delta T_b=2\times10^4$~K compared with a uniform UVB for simulation comoving box sizes $10-80h^{-1}$~cMpc and initial gas particle numbers $1024^3-2048^3$, for the standard $\Lambda$CDM model from the Sherwood-Relics simulations. Also shown is a WDM model with particle mass 2~keV. The relative change $\delta P_F(k)$ is well-converged for $k>0.005\,\mathrm{s\,km^{-1}}$ (the fundamental frequency of the $10h^{-1}$~cMpc boxes at $z=4.6$), except the $10h^{-1}$~cMpc box is required to recover $\delta P_F(k)$ at $k\gsim0.3\,\mathrm{s\,km^{-1}}$. The same relative change is found for the WDM model as for the $\Lambda$CDM model, so that the thermal suppression of power is decoupled from the suppression arising from the free-streaming of the dark matter particles. The change in power at longer wavelengths, $k<0.005\,\mathrm{s\,km^{-1}}$, is converged for the larger box sizes $40h^{-1}$~cMpc and $80h^{-1}$~cMpc.

\begin{figure*}
%\scalebox{0.5}{\includegraphics{fig01_a.eps}\includegraphics{fig01_b.eps}}
\scalebox{0.6}{\includegraphics{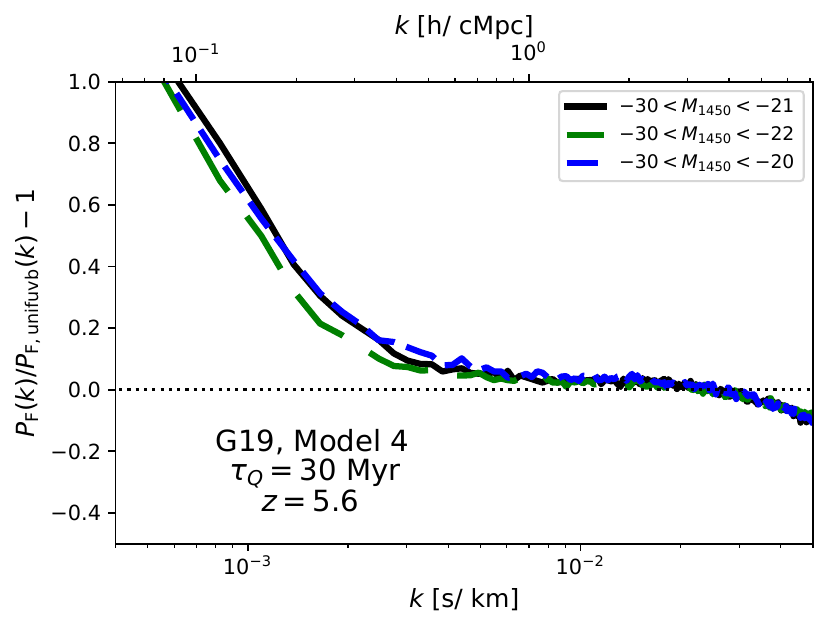}\includegraphics{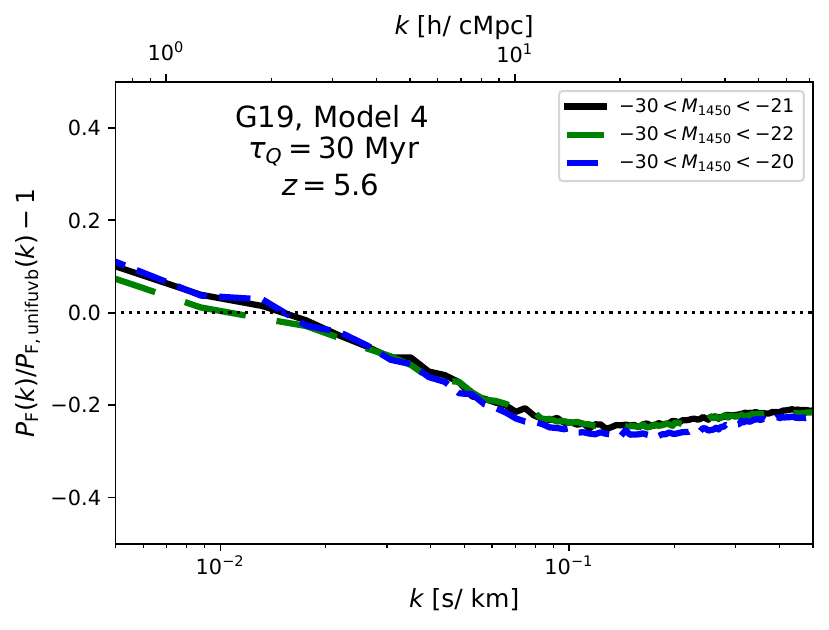}}
%\vspace{-1.5cm}
\caption{The relative difference in the \Lya\  pixel flux power spectrum at $z=5.6$ including QSOs following the G19 Model 4 QSO luminosity function, for \HeIII-region temperature boost, $\Delta T_b=2\times10^4$~K, QSO lifetime $\tau_Q=30$~Myr and the indicated maximum magnitude for sampling the QSO luminosity function. {\it Left-hand panel}:\ The box size is $160h^{-1}$~cMpc. {\it Right-hand panel}:\ The box size is $10h^{-1}$~cMpc.
}
\label{fig:dPkF_dTb_G19_Mmax}
\end{figure*}

The sensitivity of the relative difference in the \Lya\ flux power spectrum compared with a uniform UVB at $z=5.6$ to the maximum magnitude sampled from the G19 Model 4 QSO luminosity function is shown in Fig.~\ref{fig:dPkF_dTb_G19_Mmax}. For the narrower range in QSO luminosities, with $M_{1450}<-22$, the \HeIII\ filling factor is reduced to 0.18 compared with 0.22 and 0.24 for $M_{1450}<-21$ and $-20$, respectively. The reduced \HeIII\ factor results in less enhancement in the power spectrum at low wavenumbers. The suppression in power at high wavenumbers is little affected, with a slight enhanced amount of suppression for $M_{1450}<-20$.

\label{lastpage}

\end{document}